\documentclass[12pt]{article}
 \pdfoutput=1
\usepackage{jcappub}
\usepackage[top=1in, bottom=1in, left=1in, right=1in]{geometry}
\usepackage[font=footnotesize]{caption}
\usepackage{array}
\usepackage{enumerate}
\usepackage{float}
\usepackage{subfig}
\usepackage{multicol}
\usepackage{multirow}
\usepackage{epsfig}
\usepackage{graphicx}
\usepackage{pict2e}
\usepackage{color}
\usepackage{amsmath,amsfonts,amssymb}
\usepackage{bbm}
\usepackage[utf8]{inputenc}
\usepackage[english]{babel}
\usepackage{xspace}

\newcommand{\be}{\begin{equation}}
\newcommand{\ee}{\end{equation}}
\newcommand{\bea}{\begin{eqnarray}}
\newcommand{\eea}{\end{eqnarray}}

\title{Gravitational Waves in Bouncing Cosmologies from Gauge Field Production}

\begin{abstract}
 {We calculate the gravitational waves (GW) spectrum produced in various Early Universe scenarios from gauge field sources, thus generalizing earlier inflationary calculations to bouncing cosmologies. We consider generic couplings between the gauge fields and the scalar field dominating the energy density of the Universe. We analyze the requirements needed to avoid a backreaction that will spoil the background evolution. When the scalar is coupled only to  $F \tilde F$ term, the sourced GW spectrum is exponentially enhanced and parametrically the square of the vacuum fluctuations spectrum, ${\cal P}^s_T\sim ({\cal P}^v_T)^2$, giving an even bluer spectrum than the standard vacuum one. 
 When the scalar field is also coupled to $F^2$ term, the amplitude is still exponentially enhanced, but the spectrum can be arbitrarily close to scale invariant (still slightly blue), $n_T\gtrsim 0$, that is distinguishable form the slightly red inflationary one. Hence, we have a proof of concept of observable GW on CMB scales in a bouncing cosmology.}
  \end{abstract}

\author{Ido Ben-Dayan}
\affiliation{Department of Physics,
Ben-Gurion University of the Negev, P.O. Box 653, Be'er-Sheva 8410500, Israel}
\emailAdd{ido.bendayan@gmail.com}
\begin{document}

\maketitle

\section{Introduction}

Modern Cosmology is based on a myriad of observations leading to the celebrated "Concordance Model". The accuracy and precision of the model is mostly based on accurate CMB temperature anisotropies measurements carried out by telescopes, ballon and satellite experiments \cite{Ade:2015xua, Array:2015xqh, deHaan:2016qvy, vanEngelen:2014zlh, Riess:2016jrr}. 
From current observations of the temperature anisotropies we infer a scalar power spectrum with a slightly red tilt, $A_S\simeq 2.1\times10^{-9},\, n_S\simeq 0.96$ and an upper bound on the ratio between the tensor and scalar spectrum $r\equiv {\cal P}_T/{\cal P}_S|_{k_0}<0.07$. 
The ruling paradigm, 'inflation', suggests that during a period of accelerated expansion in the very early Universe, vacuum fluctuations of the elusive inflaton generated a nearly scale invariant primordial scalar power spectrum (density fluctuations) and tensor power spectrum (GW). These fluctuations are imprinted on the CMB temperature anisotropies and the CMB B-mode polarization.  

Various models of inflation successfully fit current data, though many models, including the simplest ones, are now ruled out in the absence of measuring $r$, \cite{Array:2015xqh, Ade:2015lrj}. 
Nevertheless, inflation is not problem free. Even if we ignore discussions of speculative nature, \cite{Guth:2013sya,Ijjas:2014nta}, it is clear that inflation is geodesically past incomplete \cite{Borde:2001nh} and requires a UV completion to properly address the Big Bang singularity. Thus, even though inflation pushed the Big Bang singularity much further into the past than in the Hot Big Bang cosmology, the singularity is still there, waiting for a resolution. 
Once there is some New Physics 'resolving' the Big Bang singularity, it makes sense to consider whether this New Physics discards inflation altogether, yielding a more comprehensive picture of Early Universe phenomena.

There are two possible ways of 'resolving' the singularity. First, using full fledged quantum gravity theory, to describe the singularity from which inflation or its alternatives, such as string gas cosmology ensue \cite{Brandenberger:2011et, Battefeld:2014uga}. It is fair to say that our knowledge of quantum gravity is somewhat lacking for realistic consideration. Second, using effective field theory (EFT), we can consider a 'bounce', i.e. the Universe never reaches a singularity, but rather, at some finite scale, changes its evolution from contraction to expansion. This way, gravity can be treated semi-classically, in a manner similar to inflation, and we can make testable predictions as long as the field theory is self-consistent. For a bouncing Universe, violation of the Strong Energy Condition (SEC) seems sufficient \cite{Graham:2011nb}, but in order to have an alternative to inflation, it seems a violation of the Null Energy Condition (NEC) is necessary, which is what we seek \cite{Cai:2007qw, Cai:2008qw, Lehners:2010fy}. A violation of the NEC can result in a ghost instability, gradient instability and a graceful exit problem. We shall not discuss these issues, but focus on the period of a slow contraction, where a scalar field evolves mostly with the canonical kinetic term, and that such a period can be smoothly embedded in a ghost condensate or a G-bounce model (e.g. \cite{Lehners:2010fy, Easson:2011zy, Cai:2012va}), that allow for a 'healthy' bounce without modifying our predictions.

In these EFT bouncing models, the scalar spectrum are still generated by quantum fluctuations of the 'bouncer', i.e. the field(s) that dominate the energy density of the Universe during the contraction or by some other fields in the particle spectrum. The tensor spectrum is generated by the spacetime vacuum fluctuations. In that sense, while the background evolution is very different from inflation, (exponentially fast expansion vs. slow contraction), the mechanism that generates the observed CMB spectra is similar.

An early attempt considering an alternative Early Universe Cosmology lies in the so called 'Pre Big Bang' scenario \cite{Veneziano:1991ek, Gasperini:1992em}. Based on string theory considerations, a period of slow contraction is generated when an initially weakly coupled dilaton evolves towards a strong coupling regime and gets trapped at some minimum of a potential, or evolves endlessly \cite{Gasperini:2002bn}. A nearly scale invariant power spectrum is achieved by a Kalb-Ramond axion acting as a 'curvaton' \cite{Gasperini:2002bn}. The relic graviton spectrum generated during the contraction was calculated for the first time in the Pre Big Bang scenario and the result is a very blue spectrum \cite{Gasperini:2002bn,  Gasperini:1992dp, Brustein:1995ah, Brustein:1996ut}.\footnote{A calculation of the relict gravitational radiation spectrum given some phase (de Sitter or bounce) prior to radiation domination has already been performed in the early days of the theory of cosmological perturbations in \cite{Starobinsky:1979ty}.}

The simplest, and certainly most conservative, attempt for a 'bouncing cosmology' is the so called matter bounce. If the energy density of the Universe during contraction is mostly dust, with equation of state $w\simeq 0$, then a nearly scale invariant scalar and tensor spectra are produced. It is easy to understand this result because the Mukhanov-Sasaki equation is invariant under $b\rightarrow 1-b$ for $a\sim (-\tau)^b$, where $\tau$ is conformal time. As such the 'matter bounce' and inflation are dual to one another. The problem with the matter bounce, is that it suffers from the BKL instability. During contraction, as we approach the bounce, any anisotropy will become dominant since it scales like $a^{-6}$, where $a$ is the scale factor of the FLRW metric, thus ruining the statistical isotropy of the observed Universe on large scales \cite{Battefeld:2014uga}.

In order to avoid the BKL instability, one has to consider a period of contraction with an equation of state $w>1$.
If so, the energy density of that component will scale as $a^{3(1+w)}$, and will dominate the energy density of the Universe approaching the Bounce. 
Single field slow contraction, with $w \gg1$ generates a blue spectrum for both scalar and tensor spectra. However, considering two fields, red tilted scalar power spectrum is achievable, in accord with observations \cite{Lehners:2010fy}.
Generically such bouncing models also tend to have larger non-gaussianity \cite{Lehners:2010fy,Lehners:2013cka}.

The recurring theme in all such constructions (single or multi-field) is a blue spectrum of GW, $n_T\sim 2$, compared to the inflationary $-1\ll n_T\sim -2\epsilon<0$. For a given $n_T$, the need to fulfil BBN bounds and the number of effective neutrinos $N_{eff}$, then place stringent constraints the amplitude of the GW spectrum at CMB scales making it unobservable \cite{Lehners:2010fy,Lehners:2013cka, Boyle:2003km, Buchbinder:2007ad,Chowdhury:2015cma} \footnote{In principle, given the blue spectrum, one may hope that GW experiments like Advanced LIGO and eLISA can detect it \cite{Abbott:2016blz, AmaroSeoane:2012km}. However, given that these experiments will be dominated by GW from compact objects, it is unclear whether a clear detection of a primordial signal is feasible.}. The robustness of the GW signal in the context of inflation compared to alternatives has been analyzed in \cite{Creminelli:2014wna, Geshnizjani:2014bya}.
For this reason, the fact that the GW spectra is directly linked to the energy scale of inflation ${\cal P}_T\sim H^2/M_{pl}^2$, and due to the robustness and simplicity of inflationary predictions, detection of GW on CMB scales is many times considered as a 'proof' of inflation.

The aim of this work is to find a counter example. We present a way to generate nearly scale invariant GW spectra, and potentially observable on CMB scales in the ekpyrotic slowly contracting Universe. We achieve this by considering the interaction of the bouncer with gauge fields. The interaction with the gauge fields generates sourced scalar and tensor fluctuations and gives rise to additional spectra on top of the standard vacuum analysis.
The idea that gauge fields modify the standard inflationary predictions has been extensively considered with rich phenomenology including distinct GW and scalar spectra, non-gaussianity, magnetogenesis and baryogenesis, e.g. \cite{Brustein:1999rk, Ben-Dayan:2016gxw, Barnaby:2010vf, Sorbo:2011rz, Barnaby:2011vw, Barnaby:2012tk, Barnaby:2012xt, Pajer:2013fsa, Caprini:2014mja, Mirbabayi:2014jqa, Anber:2015yca, Namba:2015gja,jiro1, jiro2,  Domcke:2016bkh}. 
One of the important results of the analyses was disentangling the connection between the energy scale of inflation and a detectable GW signal \cite{Sorbo:2011rz, Barnaby:2011vw}.  Observable GW on CMB scales via interaction with gauge fields is possible even when inflation occurs far below the GUT scale.

We apply this idea to the bouncing cosmology scenario. The analysis has many similarities with the inflationary one. In particular, the gauge field behavior is the same as in inflation. This is reminiscent of the inflation-contraction duality, that has been discussed in the past \cite{Brustein:1998kq, Wands:1998yp}. The main difference are the vacuum fluctuations of the scalar and tensor, and as a byproduct, the corresponding Green's function utilized for the analysis of the sourced fluctuations.
Our results are given in terms of background parameters and are very general. As such, the analysis applies to various FLRW cosmologies (inflation, matter bounce, ekpyrosis and others), and can also reproduce previous inflationary results. 
We limit ourselves to $U(1)$ gauge fields. It can be the Standard Model $U(1)_Y$ gauge group or some other putative $U(1)$. 

We consider two generic scenarios. First, we start with a coupling between the bouncer $\varphi$ and the gauge fields of the form $\varphi F\tilde F$. In such case, the sourced spectrum is exponentially enhanced ${\cal P}_T^s \sim e^{4\pi \xi},\, \xi>1$, and it is parametrically the square of the vacuum fluctuations spectrum ${\cal P}_T^s\sim e^{4\pi \xi}\left({\cal P}^v_T\right)^2$. Therefore, for inflation and a matter bounce, it generates a nearly scale invariant GW spectrum with $n_T^s \simeq 2n_T^v$, that is potentially observable on CMB scales. On the other hand, for a bouncing model such as ekpyrosis, the outcome is an even bluer spectrum, 
 and hence completely undetectable on CMB scales. 
In the second model, we consider a more generic coupling also to the Maxwell term $(-\tau)^{-n}\left(F^2-\gamma F\tilde F\right)$. We shall see that there is a duality relating spectra ${\cal P}_T(n,\gamma ;\,n\geq-1/2)={\cal P}_T \left(-1-n, -\gamma n/(1+n);\, n\leq -1/2\right)$, 
so we can always choose $n$ that gives a weakly coupled $U(1)$.
In this case we still get the exponential enhancement ${\cal P}_T^s\sim e^{4\pi \xi}$ but also a tilt controlled by $n$, such that it can be arbitrarily close to a scale invariant spectrum, but slightly blue $n_T\gtrsim 0$, well within the allowed parameter space of $n_T\leq 0.28-0.63$ depending on the assumptions we will specify below.
Hence, such a spectrum is observable on CMB scales with $r\sim 10^{-4}-10^{-2}$, which can hopefully be observed within the next decade. Contrary to the vacuum fluctuations, only one chirality of the sourced spectrum is enhanced, and therefore such a spectrum is distinguishable from the vacuum one \cite{Sorbo:2011rz}. More importantly, the spectrum is nearly scale invariant, but slightly blue! Unlike the generic inflationary predictions that suggest a slightly red spectrum\footnote{See however \cite{Mukohyama:2014gba}.}. Thus, the measurement of the tilt of the gravitational power spectrum will be a decisive factor in favoring inflation or a bouncing cosmology.

The results presented here are therefore a proof of concept. It remains to be checked whether the mechanism presented here can be merged with a two field scenario that provides for the entropic mechanism of generating the nearly scale invariant scalar fluctuations \cite{Lehners:2010fy}, and whether the mechanism in its various forms fulfils non-gaussianity bounds \cite{Ade:2015ava}. Alternatively, one may consider the gauge field production as a source of the scalar spectrum as well, and how such spectrum sails through the bounce. We are currently analyzing this possibility.

The paper is organized as follows. In the next section we discuss the general setup and review the basic dynamics of a slowly contracting Universe. In section \ref{sec:constraints} we review the current and near future constraints on the tilt of the GW spectrum. In section \ref{sec:inclusion} we discuss the inclusion of gauge fields in a bouncing cosmology. In section \ref{sec:backreaction} we analyze the conditions that the gauge fields backreaction do not spoil the background dynamics. In section \ref{sec:formal-sol} we present the formal setup for calculation of the power spectrum, and generalize the Green's function method to general flat FLRW cosmologies. In section \ref{sec:model1} we evaluate the GW spectrum for model I. In section \ref{sec:model2} we evaluate the GW spectrum for model II. In section \ref{sec:observations} we compare our results with observations and backreaction constraints. We then provide some final remarks in \ref{sec:final}. 

\section{Setup}
\label{sec:setup}
We shall consider two scenarios. For concreteness, we shall many times take ekpyrosis as the explicit model in calculations, but wherever possible, we will note that the analysis holds for other scenarios as well. Model I: The bouncer is coupled to $F\tilde F$. Model II: A time dependent coupling, that can be written in terms of the bouncer to both $F^2$ and $F \tilde F$:
\begin{eqnarray}
\label{L1}
  \mathcal{S}_1 =  \int d^4 x \sqrt{-g} \left[ \frac{M_p^2}{2} \, R -\frac{1}{2}(\partial \varphi_1)^2 - V(\varphi_1) - \frac{1}{4}F^{\mu\nu}F_{\mu\nu} - \frac{\alpha}{4 f} \varphi_1  \, \tilde{F}^{\mu\nu} \, F_{\mu\nu} \right]\\
  \label{L2}
  \mathcal{S}_2 =  \int d^4 x \sqrt{-g} \left[ \frac{M_p^2}{2} \, R -\frac{1}{2}(\partial \varphi_1)^2 - V(\varphi_1) - I^2(\tau)\left\{\frac{1}{4}F^{\mu\nu}F_{\mu\nu} - \frac{\gamma}{4} \, \tilde{F}^{\mu\nu} \, F_{\mu\nu}\right\} \right]
\end{eqnarray}
Note that $\tilde F^{\mu \nu}=\frac{1}{2}\epsilon^{\mu \nu \rho \sigma}F_{\rho \sigma}$, $\varphi_1$ is the field driving the slow contraction, and $\tau$ denotes conformal time. We shall later write down the form of $I(\tau)\equiv (-\tau)^{-n}$ in terms of the $\varphi_1$ field.
 For arbitrary $U(1)$, both positive and negative values of $n$ are allowed, since we can assume that there are no fields charged under this $U(1)$, and it is basically a free theory. For the SM $U(1)_Y$ or $U(1)_{em}$ we have to limit ourselves to $n\leq0$ to avoid strong coupling of the theory at very early times such as the onset of contraction. 
 Regardless, we will use notation such as $\vec E, \vec B$, throughout the paper.   In our analysis we shall see that the equations for the mode functions are invariant under $n\rightarrow -1-n, \gamma \rightarrow -\gamma n/(1+n)$. The outcome being ${\cal P}_T(n,\gamma;\,n\geq-1/2)={\cal P}_T \left(-1-n,-\gamma n/(1+n);\, n\leq -1/2 \right)$, making the strong vs. weak coupling discussion rather immaterial. 

 In this work, we limit ourselves to the tensor power spectrum. For a full account of Early Universe phenomena, one has to generate a viable scalar power spectrum, show how the tensor and scalar spectra pass through the bounce and check that the model does not violate non-gaussianity bounds.
 Given these reservations, the interesting question still remains, can we generate a detectable tensor spectrum on CMB scales?
Some optimism may be drawn from a recent analysis of a similar model generating primordial magnetic fields during the contracting phase. In that analysis the magnetic field power spectrum survives the bounce \cite{Chowdhury:2016aet}.

\subsection{Contracting Background Evolution}
There are several reviews on bouncing cosmologies \cite{Lehners:2010fy, Battefeld:2014uga, Brandenberger:2016vhg, Gasperini:2002bn, Cai:2014bea} and we refer the reader interested in more details about bouncing cosmologies to these reviews and references therein. Here we simply review the most basic background equations, and give some general relevant remarks. In the absence of gauge fields, an exact 'scaling solution' of the background equations is given by:
\be
\label{potential}
V=-V_1e^{-\sqrt{2/p}\varphi_1}, \quad a(t)\sim (-t)^p, \quad -\infty\leq t\leq 0
\ee
$\varphi_1$ is written in Planck units and $p>0$ is dimensionless. Power law inflation corresponds to $p \gg 1, V_1<0$, ekpyrosis to $p \ll 1, V_1>0$ and the matter bounce to $p=2/3, V_1>0$.
The generalization to several fields is:
\bea
a=(-t)^p, \quad p\ll1, \quad p=\sum_i\frac{2}{c_i^2}, \quad H=\frac{p}{t}\\
\label{eq:field}
\varphi_i=\frac{2}{c_i}M_{pl}\ln \left(-\sqrt{\frac{V_i}{2/c_i^2(1-3p)M_{pl}^2}}t\right), \,
  \dot{\varphi_i}=\frac{2}{c_i t} \, , \quad \ddot{\varphi_i}=-\frac{2}{c_i t^2} ,\\
 V=-\sum_i V_ie^{-c_i\varphi_i}=-\frac{p(1-3p)}{t^2}, \quad V_i,c_i>0 \, \forall i
\eea
In the context of ekpyrosis, one has "fast-roll" conditions with $\epsilon=-\dot{H}/H^2=1/p \gg1$, and $\eta \gg 1$. 
In more realistic scenarios, by properly choosing $c=c(\varphi_1)$, $\epsilon$ is not constant and the universe exits from ekpyrosis. One also gets a slightly red scalar spectral index in accord with observations through the entropic mechanism \cite{Lehners:2010fy}. 
Additionally, in inflation $H$ is nearly constant while $a$ grows exponentially. In the slow contraction scenario, $a$ is nearly constant and slowly approaches Minkowski space, while $H$ is varying.
Hence, the Hubble friction term is negligible compared to the terms existing in flat space as well, as long as $\epsilon \gg 1$.
An interesting private case is that of a matter bounce, that corresponds to $p=2/3$. In this contracting phase, a scale invariant spectrum from adiabatic fluctuations is automatically generated for both scalars and tensors. However, as is well known such a scenario suffers from the anisotropy instability \cite{Battefeld:2014uga}. The main reason for requiring a slow contraction with $p\ll1$ is exactly to overcome this instability.

After the ekpyrotic phase, there is a kinetic energy domination contracting phase $a\sim (-t)^{1/3},\, t<0$, the Bounce $t=0$, kinetic energy domination expanding phase $a\sim t^{1/3},\, t>0$ followed by the standard radiation domination phase of the Hot Big Bang \cite{Lehners:2010fy, Boyle:2003km}.


We will also use conformal time $-\infty<\tau\leq 0$ with $-\tau=(-t)^{1-p}/(1-p)$, and plug it in the background solution to \eqref{eqA}.
\bea
a(t)\sim(-t)^p=(-(1-p)\tau)^{p/(1-p)}\, ,\quad \mathcal{H}=\frac{p}{(1-p)(-\tau)}\\
\varphi_i'=\frac{2}{c_i(1-p)\tau}\, , \quad \varphi_i''=-\frac{2}{c_i(1-p)\tau^2}\, , V=-\frac{p(1-3p)}{\left(-(1-p)\tau\right)^{2/(1-p)}}
\eea

 We normalize the scale factor, such that it is unity at the end of ekpyrosis, $a(\tau)=(-\tau/\tau_{end})^b$, and $b\equiv p/(1-p)$. In ekpyrosis, $p\ll1$, so $b\simeq p$. \footnote{ Inflation (exact de Sitter) corresponds to $b=-1, p\rightarrow \infty$ and the matter bounce to $p=2/3,\,b=2$. In such cases we have to keep all powers of $b$ or $p$ in the analysis.}

\subsection{Tensor Power Spectrum from Vacuum Fluctuations}
Let us quickly review the vacuum fluctuations analysis. 
 Define $\hat Q_k=a \hat h_k$, where $\hat h_k$ is the tensor perturbation.  The EOM for the tensor perturbation in the presence of a source is given by:
 \begin{equation}
\left[ \partial_\tau^2 + \left( k^2 - \frac{a''}{a} \right) \right]  Q_\lambda \left( \tau ,\, \vec{k} \right) = J_\lambda \left( \tau ,\, \vec{k} \right)
\label{hc-eq-formal}
\end{equation}

Decomposing into vacuum and sourced fluctuations:
\begin{align}
Q_k \left( \tau \right)&=Q^v_{\vec k}(\tau)+Q^s_{ \vec k}(\tau)\\
Q_k^v\left( \tau \right)&=b(\vec k)f_k(\tau)+b^{\dagger}(-\vec k)f^*_k(\tau),\quad \left[b(\vec k),b^{\dagger}(\vec k')\right]=\delta^{(3)}(\vec k-\vec k')
\end{align}
The calculation in the presence of vacuum fluctuations only, is given by $J_{\lambda}\equiv 0$. The mode functions then obey:  
\be
f_k''+\left(k^2-\frac{a''}{a}\right)f_k=0
\ee
With initial conditions reproducing Minkowski space at $\tau\rightarrow -\infty$, the tensor perturbation reads:
\be
h=\frac{\sqrt{2\pi}}{M_{pl}}\frac{\sqrt{-\tau}}{a}H^{(1)}_{1/2-b}(-k\tau)
\ee
Since $a=(-\tau/\tau_{end})^b, \quad b\equiv p/(1-p)$, Inflation means $b=-1$, ekpyrosis means $0<b \ll1$ and matter bounce $b=2$. Notice that for $a\sim (-\tau)^b$ the equation is invariant under the transformation $b\rightarrow 1-b$, therefore we have duality between different values of $b$, this is why both inflation and the matter bounce give similar spectra. We normalized $a$ to be unity at the end of the slow contraction phase, $\tau_{end}$.
Hence the spectrum from vacuum fluctuations, outside the horizon is:
\be
{\cal P}_T=\frac{k^3}{2\pi^2} \Big \langle hh \Big \rangle.
\ee
For inflation (and matter bounce) we have the well known (nearly) scale invariant result: 
\be
{\cal P}_T=\frac{4 H^2}{\pi^2 M_{pl}^2}.
\ee
In ekpyrosis, a similar calculation has been carried out in \cite{Boyle:2003km}. 
The index of the Hankel function is about $1/2$ instead of $3/2$ so we get:
\bea
\label{PTv}
\Big \langle hh \Big \rangle_{ekp.}=2\times \frac{4^{1-b}\Gamma(1/2-b)^2}{\pi M_{pl}^2k_e^{2b}}k^{-1+2b}\, \Rightarrow {\cal P}_T=2\times \frac{4^{1-b}\Gamma(1/2-b)^2}{2 \pi^3 M_{pl}^2k_e^{2b}}k^{2+2b} \, \rightarrow_{b\rightarrow0}\frac{4k^{2}}{\pi^2M_{pl}^2}\cr
\eea
where $k_e=\tau_{end}^{-1}$ is from the normalization of the scale factor at the end of the slow contraction phase. ${\cal P}_T\sim k^{2+2b}$ is a highly blue spectrum.

\section{Constraints on the GW Spectrum}
\label{sec:constraints}
Consider the standard primordial GW and scalar spectra parameterization at some pivot scale $k_0$:
\be
{\cal P}_T=A_T(k_0) \left(\frac{k}{k_0}\right)^{n_T},\quad
{\cal P}_S=A_S(k_0) \left(\frac{k}{k_0}\right)^{n_S-1}, \quad r\equiv \frac{A_T}{A_S}|_{k_0}
\ee
Best current bounds on $r$ from CMB polarization experiments are $r_{0.05}\leq0.07$.
We can use the recent analysis by \cite{Meerburg:2015zua} as to bounds on the allowed tensor tilt. Since we do not derive a scalar spectrum, we shall simply assume that such a spectrum is somehow generated without spoiling our predictions for the GW spectrum.
Assuming that $n_T$ is constant, the bound on the allowed $n_T$ and tensor to scalar ratio $r$ is derived from the allowed number of relativistic species during recombination from CMB measurements \cite{Ade:2015xua}:
\be
N_{eff}=3.15\pm0.23
\ee
The value predicted by known physics is $N_{eff}=3.046$ in perfect accord with PLANCK. CMB Stage $3$ experiments are expected to improve the measurement to $\Delta N_{eff}=0.06$ and CMB Stage $4$ to  $\Delta N_{eff}=0.02$, as well as improving the accuracy of measuring $r$ to $\Delta r\sim 10^{-4}$ \cite{Abazajian:2013oma, Wu:2014hta, joel}.

Given that GW are obviously relativistic degrees of freedom, though subdominant compared to the energy density of the Universe $\rho_{GW}\ll \rho_{total}$, the constraint from $N_{eff}$ reads \cite{Meerburg:2015zua}:
\be
N_{eff} \simeq 3.046+\left(3.046+\frac{8}{7}\left(\frac{11}{4}\right)^{4/3}\right)\frac{A_S r}{24 n_t}\left(\frac{k_{UV}}{k_0}\right)^{n_T}
\ee
Obviously, $N_{eff}$ is exponentially sensitive to $n_T$, especially since $k_{UV}/k_0$ is the range of wavenumbers that have experienced the slow contraction/inflation. We consider three ranges of wave numbers:
(a) Wavenumbers up to the LIGO scales, since we now have an actual observation, $k_{UV}/k_0=10^{20}$. (b) $60$ e-folds of inflation/contraction $k_{UV}/k_0=10^{24}$ and (c) The Planck scale $k_{UV}/k_0=10^{30}$. 
Substituting from the above calculation $n_T=2$ and $A_s=2.1\times 10^{-9}$ requires $r<2\times 10^{-31}$ to be within two standard deviations of the present measurement. Obviously, this is undetectable in any foreseeable future. 
If the current bound on $r$ is actually measured, $r=0.07,\,k_{UV}/k_0=10^{24}$ stage 4 will constrain the tilt to be $n_T\leq0.35$, at $2$-sigma, while the current PLANCK constraints on $N_{eff}$ and $r=10^{-4},\,k_{UV}/k_0=10^{20}$ allow $n_T \leq 0.63$. Finally, one can even consider a spectrum all the way up to the Planck scale $k_{UV}/k_0=10^{30}$ and $r=0.07$ such that Stage $4$ CMB experiment will put the upper bound $n_T\leq 0.28$. Hence, assuming $r\geq 10^{-4}$, we can expect at most a factor of two improvement of the bound on $n_T$ from the most lenient assumptions with today's data, to the most conservative assumptions and with future data.

\begin{figure}
\centerline{
\includegraphics{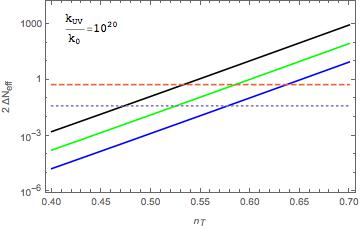}
}
\caption{Logplot of $2 \Delta N_{eff}$ as a function of $n_T$ for various values of $r$, and for $k_{UV}/k_0=10^{20}$. Solid black $r=10^{-2}$, solid green $r=10^{-3}$ and solid blue $r=10^{-4}$. The horizontal dashed line corresponds to present PLANCK constraints $2 \Delta N_{eff}=0.46$ and the horizontal dotted line corresponds to stage 4 CMB experiment $2 \Delta N_{eff}=0.04$.
}
\label{fig:neff1}
\centerline{
\includegraphics{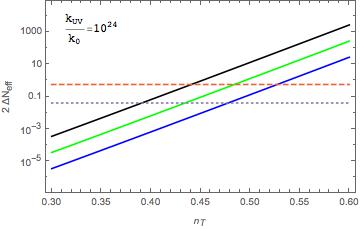}
}
\caption{Same as figure $1$ but with $k_{UV}/k_0=10^{24}.$
}
\label{fig:neff2}
\end{figure}

\begin{figure}
\centerline{
\includegraphics{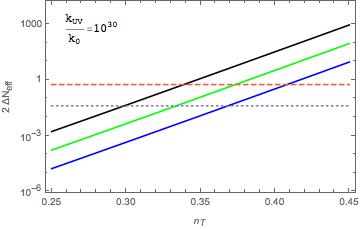}
}
\caption{Same as figure $1$ but with $k_{UV}/k_0=10^{30}.$
}
\label{fig:neff3}
\end{figure}

In figure \ref{fig:neff1}, assuming $k_{UV}/k_0=10^{20}$, we show the current and future constraints on $n_T$ for $r=10^{-4},10^{-3},10^{-2}$, the latter two certainly within observational reach. In figures \ref{fig:neff2},\ref{fig:neff3}, we repeat the analysis with $k_{UV}/k_0=10^{24},k_{UV}/k_0=10^{30}$, respectively. 
Let us note that the actual bound on $n_T$ is probably more relaxed. The reason is that the smallest modes (with the most power) will re-enter the horizon during kinetic energy domination and will change their tilt $n_T\rightarrow n_T-1$. We expect that a careful analysis will relax the bound from $N_{eff}$ a bit.  



\section{Inclusion of Gauge Fields}
\label{sec:inclusion}
As mentioned in the Introduction, the effect of gauge fields on inflation has been extensively studied, and results in amplification of GW spectrum. If one wishes to have any amplification of GW through this mechanism in a bouncing Universe,
the crucial question is whether gauge field production is enhanced. We now review the basic relevant equations in the case of inflation and a contracting Universe. Decomposing the vector potential into modes:
\begin{equation}
  \vec{A}(\tau,{\vec x}) = \sum_{\lambda=\pm} \int \frac{d^3k}{(2\pi)^{3/2}} \left[ \vec{\epsilon}_\lambda({\vec k}) a_{\lambda}({\vec k}) A_\lambda(\tau,{\vec k}) e^{i {\vec k}\cdot {\vec x}} + \mathrm{h.c.}   \right]
\label{decomposition}
\end{equation}
 The annihilation and creation operators obey:
\begin{equation}
  \left[a_{\lambda}({\vec k}), a_{\lambda'}^{\dagger}({\vec k'})\right] = \delta_{\lambda\lambda'}\delta^{(3)}({\vec k}-{\vec k'})
\label{ladder}
\end{equation}
Here $\vec{\epsilon}_\lambda$ are circular polarization vectors satisfying  $\vec{k}\cdot \vec{\epsilon}_{\pm} \left( \vec{k} \right) = 0$, 
$\vec{k} \times \vec{\epsilon}_{\pm} \left( \vec{k} \right) = \mp i k \vec{\epsilon}_{\pm} \left( \vec{k} \right)$,
$\vec{\epsilon}_\pm \left( \vec{-k} \right) = \vec{\epsilon}_\pm \left( \vec{k} \right)^*$, and normalized according to $\vec{\epsilon}_\lambda \left( \vec{k} \right)^* 
\cdot \vec{\epsilon}_{\lambda'} \left( \vec{k} \right) = \delta_{\lambda \lambda'}$.  
The annihilation and creation operators of the gauge field commute with the operators of the tensor and scalar fluctuations:
\be
\label{abcommute}
 \left[b({\vec k}), a_{\lambda'}^{\dagger}({\vec k'})\right] =\left[b({\vec k}), a_{\lambda'}({\vec k'})\right]=0
\ee 

In Model I, the EOM for the gauge field is \cite{Barnaby:2011vw}:
\begin{equation}
\vec{A}'' - \nabla^2 \vec{A} - \frac{\alpha}{f} \, \varphi_1' \, \vec{\nabla} \times \vec{A} = 0.
\label{eqA}
\end{equation} 

\subsection{Inflation Model I}

In inflationary background and going to Fourier space, the mode functions obey: 
\begin{equation}
\label{Amode}
  \left[ \frac{\partial^2}{\partial\tau^2} + k^2 \pm \frac{2 k \xi}{\tau} \right] A_{\pm}(\tau,k) = 0, \hspace{5mm} \xi \equiv \frac{\alpha \dot{\varphi_1}}{2 f H}
\end{equation}
Notice that $\xi$ is a slow-roll quantity, $\xi=\frac{\alpha M_{pl}}{f}\sqrt{2\epsilon}$. So for $\xi>1, \, M_{pl}/f$ has to over come the slow-roll suppression.
The crucial point is that if $\xi>1$ then the amplitude of the $+$ polarization is exponentially enhanced by a factor $e^{\pi \xi}$ when exiting the horizon:
\begin{equation}
\label{mode_soln}
  A_{+}(\tau,k) \cong \frac{1}{\sqrt{2k}} \left(\frac{-k\tau}{2\xi}\right)^{1/4} e^{\pi \xi - 2\sqrt{-2 \xi k \tau}}
\end{equation}


\subsection{Ekpyrosis Model I}
In a contracting universe equations \eqref{eqA},\eqref{Amode},\eqref{mode_soln} are the same, only now 
\bea
 \xi=\alpha \frac{M_{pl}}{f}\frac{\sqrt{p}}{\sqrt{2}(1-p)}\sim \alpha \frac{M_{pl}}{f\sqrt{2\epsilon}}
\eea
The main point is that we have the same behavior. 
Here $\epsilon \gg1$ so again we need $f\ll M_{pl}$ to beat the "fast-roll" suppression, exactly like inflation. Hence, also in a contracting universe gauge fields will be exponentially enhanced when $\xi >1$. 
Specifically, we see that in the matter bounce $\xi=\frac{\sqrt{3}\alpha M_{pl}}{f}$, so in this scenario $\xi \gg 1$ naturally, unless one fine-tunes $\alpha$ to be small and/or pushes $f\rightarrow M_{pl}$. 

Given that the equation for the mode functions is identical to that of inflation only with a different definition of $\xi$, the solutions and the range of validity will be similar to inflation, i.e.:
\begin{equation}
\label{coulomb}
  A_{+}(\tau,k) = \frac{1}{\sqrt{2k}} \left[ G_0(\xi,-k\tau) + i F_0(\xi,-k\tau)  \right]
\end{equation}
where $F_0,G_0$ are the so-called Coulomb wave functions.
The production of gauge fluctuations is interesting in the region of phase space 
$(8 \xi)^{-1} \lesssim -k\tau \lesssim 2\xi$ of phase space.  
Hence, we get simple representation of the modes:
\begin{eqnarray}
  A_{+}(\tau,k) &\cong& \frac{1}{\sqrt{2k}} \left(\frac{-k\tau}{2\xi}\right)^{1/4} e^{\pi \xi - 2\sqrt{-2\xi k\tau}} 
\nonumber\\
  A_{+}'(\tau,k) &\cong& \sqrt{\frac{2k\xi}{-\tau}} A_{+}(\tau,k)      
  \label{lorenzo}   
\end{eqnarray}
%
\subsection{Gauge Fields Model II}

In model II, we have time-dependent or field dependent coupling for both the $F^2$ term and the $F\tilde F$ term. This way, one still gets that $e^{\pi \xi}$ amplitude enhancement of the $+$ helicity on the one hand, but also a controlled $-k\tau$ dependence depending on the power $n$.
Recall that:
\be
\mathcal{S}_2 =  \int d^4 x \sqrt{-g} \left[ \frac{M_p^2}{2} \, R -\frac{1}{2}(\partial \varphi_1)^2 - V(\varphi_1) - I^2(\tau)\left\{\frac{1}{4}F^{\mu\nu}F_{\mu\nu} - \frac{\gamma}{4} \, \tilde{F}^{\mu\nu} \, F_{\mu\nu}\right\} \right]
\ee

with $\gamma>0$ without loss of generality. Let us assume that $I(\tau)\sim(-\tau)^{-n}$ and write it in terms of the bouncer field:
\be
\label{eq:Itau}
I(\tau)=(-\tau)^{-n}\equiv a_2^ne^{-n\varphi_1/a_1},
\ee
 where $a_1=(1-p)M_{pl}/\sqrt{2p}\sim M_{pl}/\sqrt{2p}$, and $a_2$ can be read off from equating $\varphi_1\equiv a_1\ln(-a_2\tau)$ with equation \eqref{eq:field}. One can try other functions of $\varphi_1$, for example $I\sim \varphi_1^n$, which will add logarithmic corrections to the behavior discussed below, (for example shifting the time of horizon exit of modes), but it will not change the qualitative behavior. 
Following \cite{Caprini:2014mja}, we can define $\tilde A=I\,A$, and quantize it.
 In the Coulomb gauge $A_0=\partial^i\,A_i=0$, 
 for the canonically normalized field, the Lagrangian becomes
\begin{equation}
{\cal L}=\frac{1}{2}\,\tilde{A}_i'{}^2-\frac{1}{2}\left(\nabla \tilde{A}_i\right)^2+\frac{1}{2}\,\frac{I''}{I}\,\tilde{A}_i^2-\gamma\,\frac{I'}{I}\,\epsilon_{ijk}\,\tilde{A}_i\,\partial_j\,\tilde{A}_k\,.
\end{equation}
Notice that the above lagrangian is invariant under:
\be
 \label{eq:duality}
 n\rightarrow -1-n, \quad \gamma \rightarrow -\gamma \frac{n}{1+n},\quad \Rightarrow \quad \xi\equiv-n\gamma  \rightarrow \xi.
\ee
Quantization of the gauge field on helicity modes $\tilde{A}_\lambda$ with $\lambda=\pm 1$ is realized by defining
\begin{equation}\label{helicity}
\hat{A}_i(\tau,\vec x)=\sum_{\lambda=\pm} \int\frac{d^3k}{(2\pi)^{3/2}}\,\epsilon_i^\lambda(\vec k)\,e^{i \vec k \cdot \vec x}
\left(\tilde{A}_\lambda(\vec k,\,\tau)\,\hat{a}_\lambda(\vec k)+\tilde{A}^*_\lambda(-\vec k,\,\tau)\,\hat{a}^\dagger_\lambda(-\vec k)\right)\,
\end{equation}
where $\epsilon_i^\lambda(\vec k)$ is the polarization vector and where $\tilde{A}_\lambda$ satisfies
\begin{equation}\label{eqalambda}
\tilde{A}_\lambda''+\left(k^2+2\,\lambda\,\xi\,\frac{k}{\tau}-\frac{n\,\left(n+1\right)}{\tau^2}\right)\,\tilde{A}_\lambda=0\,.
\end{equation}
 The invariance of the lagrangian is carried onto the above equation. The equation is invariant under $n\rightarrow -1-n,\gamma \rightarrow -\gamma n/(1+n)$ (It leaves $\xi$ invariant and does not change the last term in the equation). This symmetry will generate "mirror" spectra of gauge fields and GW around $n=-1/2$. These mirror spectra are connected by the symmetry transformation. The solution of \eqref{eqalambda} with the boundary condition of the Bunch-Davies vacuum as $\tau \rightarrow -\infty$ are the Coulomb wave functions:
\be
\tilde A=\frac{1}{\sqrt{2k}}\left(G_{-n-1}(\xi,-k \tau)+i F_{-n-1}(\xi,-k\tau)\right) 
\ee

Equation \eqref{eqalambda} has three regimes. The first one, well inside the horizon $-k\tau \gg1$, the photons are in the Bunch-Davies vacuum. The second regime corresponds to horizon exit with $1\gg-k\tau \gg 1/\xi$, where:

\begin{align}\label{aapprox}
\tilde{A}(k,\,\tau)\simeq \sqrt{-\frac{2\,\tau}{\pi}}\,e^{\pi\xi}\,K_{-1+2|n|}\left(\sqrt{-8\,\xi\,k\,\tau}\right)\,.
\end{align}
Both of these regimes were also present in model I, as expected. In model II however, on top of the exponential enhancement during horizon exit $\sim e^{\pi \xi}$, the tail is modified, and we have an additional regime, for $-k\,\tau\ll 1/\xi$:
\begin{align}\label{aapproxsm}
\tilde{A}(k,\,\tau)\simeq \sqrt{-\frac{\tau}{2\,\pi}}\,e^{\pi\xi}\,\Gamma\left(|2n+1|\right)\,\left|2\,\xi\,k\,\tau\right|^{-|n+1/2|}\,. 
\end{align}

To summarize, in a contracting Universe, enhanced gauge field production occurs due to the interaction between the bouncer and the gauge fields, in a similar manner to inflation, and under similar terms (fast-roll vs. slow-roll).
We can therefore analyze the potential backreaction of this phenomenon on the background evolution and calculate the resulting spectra due to the interaction with the gauge fields. Throughout the rest of the paper, this will be called 'sourced fluctuations' since the gauge fields are a source term in the EOM of the GW and curvature perturbation. 

\section{Backreaction}
\label{sec:backreaction}
\subsection{Model I}
The behavior of the modes and their effect on the background and perturbations evolution is where we will see a qualitative difference between a contracting universe and inflation. We evaluate the effect of the gauge fields on the background evolution by using the mean field approximation:
\begin{eqnarray}
  &&  \ddot{\varphi_1} + 3 H \dot{\varphi_1} + V'(\varphi_1) = \frac{\alpha}{f} \langle \vec{E}\cdot \vec{B} \rangle \label{mean1} \\
  && 3 H^2 = \frac{1}{M_p^2} \left[ \frac{1}{2}\dot{\varphi_1}^2 + V(\varphi_1) + \frac{1}{2} \langle \vec{E}^2 + \vec{B}^2 \rangle \right] \label{mean2}
\end{eqnarray}
with
\begin{eqnarray}
\langle \vec{E} \cdot \vec{B} \rangle &=& - \frac{1}{4 \pi^2 a^4} \int d k \, k^3 \, \frac{d}{d \tau} \vert A_+ \vert^2 \nonumber\\
\frac{1}{2}\langle \vec{E}^2+\vec{B}^2 \rangle &=& \frac{1}{4 \pi^2 a^4} \int d k \,  k^2 \left[ \vert A_+' \vert^2 + k^2 \vert A_+ \vert^2 \right]
\label{integrals-mean}
\end{eqnarray}
We use the Coulomb gauge $\hat A_0=0$ and the 'electric' and 'magnetic' fields are:
\be
\hat E_i=-\frac{1}{a^2}\hat A_i',\quad \hat B_i=\frac{1}{a^2}\epsilon_{ijk}\partial_j\hat A_k
\ee

Using \eqref{lorenzo}, in the inflationary scenario, the above integrals give the time dependence $k^4 \sim \tau^{-4}$ contribution, which in exact dS, $a=-1/H\tau$ is cancelled with the $a^{-4}$ factor, so we get a constant in \eqref{mean1}, and in realistic inflation a slow-roll suppressed evolution.
In ekpyrosis, as we have shown, the integrals are identical, so $k^4 \sim \tau^{-4}$. However, $a^{-4}\sim \tau^{-4p/(1-p)}$ so the mean field value depends on time with $\langle \vec{E} \cdot \vec{B} \rangle \sim \tau^{-4/(1-p)}\sim \tau^{-4},$ with fast-roll $\propto p\ln \tau$ corrections.
Evaluating the mean fields in the relevant range $(8 \xi)^{-1} \lesssim -k\tau \lesssim 2\xi$ gives:
\bea
\langle \vec{E} \cdot \vec{B} \rangle &\simeq&\frac{0.0082}{4 \pi^2 a^4}\frac{e^{2 \pi \xi}}{\xi^4\tau^4}=2.1\times 10^{-4}\frac{e^{2\pi \xi}}{\xi^4\tau^{4/(1-p)}}=2.1\times 10^{-4}\frac{e^{2\pi \xi}}{(\xi t)^4}\\
\frac{1}{2}\langle \vec{E}^2+\vec{B}^2 \rangle &\simeq& \frac{0.00014}{4 \pi^2 a^4}\frac{e^{2 \pi \xi}}{\xi^3\tau^4}=1.4 \times 10^{-4}\frac{e^{2\pi \xi}}{\xi^4\tau^{4/(1-p)}}=1.4 \times 10^{-4}\frac{e^{2\pi \xi}}{\xi^3 t^{4}}
\eea
So unlike inflation, here the mean field evolves with time. Let us reinsert the Hubble parameter, to find the bounds necessary for a negligible backreaction, only it is useful to bear in mind that unlike inflation, $H$ is evolving, $H=p/t$.
In ekpyrosis, the friction term $3H\dot{\varphi_1}$ is negligible, and for a negligible backreaction we require:
\be
|\frac{\alpha}{f}\langle \vec{E} \cdot \vec{B} \rangle| \ll |\ddot{\varphi}|\sim |V'|, \quad  \frac{1}{2}\langle \vec{E}^2+\vec{B}^2 \rangle \ll3H^2
\ee
Substituting the time-dependent bounds and $H=p/t$ gives:
\bea
|\frac{\alpha}{f}2.4\times 10^{-4}\frac{e^{2\pi \xi}H^4}{(\xi p)^4}|\ll \frac{2 M_{pl}H^2}{c p^2} \Rightarrow \quad \frac{H}{M_{pl}}\ll 69(p \xi)^{3/2}e^{-\pi \xi}\\ 
|1.4 \times 10^{-4}\frac{e^{2\pi \xi}H^4}{\xi^3 p^{4}}| \ll 3 M_{pl}^2H^2 \Rightarrow \quad \frac{H}{M_{pl}} \ll 146 p^2 \xi^{3/2}e^{-\pi \xi}
\eea
 Since the maximal Hubble parameter is at the end of ekpyrosis, $H=H_{end}$ in the above inequalities. In ekpyrosis, since $p \ll 1$, the second inequality is a more stringent bound parametrically due to the higher power of $p$. 
 Notice that the bounds are valid also for $p\sim 1$ as long as $|\frac{\alpha}{f}\langle \vec{E} \cdot \vec{B} \rangle| \ll |\ddot{\varphi_1}|$, hence the same bound can be used in various cosmologies. 
 It is clear that there is ample parameter space where the bounds are fulfilled and enhancement of gauge field production and GW occurs simultaneously.

In principle, since $H$ continues to grow until the bounce, the gauge fields will backreact on the background evolution also during kinetic energy domination if the gauge field production does not shut down. We assume that it shuts down at the end of ekpyrosis, during the contracting kinetic phase. This is plausible since in the two field scenario, crucial to the entropic mechanism of scalar, red tilted spectrum, the classical trajectory in field space turns \cite{Lehners:2010fy}, so we can envision a lagrangian with $\varphi_1,\varphi_2$, such that after the turn, $\varphi_1$ time-dependence is highly suppressed, so the gauge field production stops.
The analysis of such a scenario is beyond the scope of this paper. In any case, we will show that the gauge fields source GW production and that the GW production stops once kinetic energy domination starts.

\subsection{Model II}

In model II, evaluation of the backreaction proceeds as follows. For the Friedmann equation, we still require:
\be
\frac{1}{2}\langle \vec{\tilde E}^2+\vec{\tilde B}^2 \rangle \ll 3M_{pl}^2H^2
\ee
The modes with $1\gg-k\tau \gg 1/\xi$ give similar contribution the one calculated for model I. Therefore, we only have to check what new bound the new asymptotic behavior due to $n\neq0$ gives:
 
 \begin{align}
\frac{1}{2}\langle \vec{\tilde E}^2+\vec{\tilde B}^2 \rangle \simeq & \frac{1}{4\pi^2} \frac{  \left((n-2) n^2 \xi ^2+n-1\right) \Gamma
   (2 n+1)^2}{2^{3+2 n}\pi  (n-2) (1-n)}\frac{e^{2 \pi  \xi }}{ \xi ^5 t^4} ,& -\frac{1}{2}<n<1 \\
 \frac{1}{2}\langle \vec{\tilde E}^2+\vec{\tilde B}^2 \rangle \simeq & \frac{1}{4\pi^2}  \frac{ \left((n+1)^2 (n+3) \xi ^2+n+2\right)
   \Gamma (-2 n-1)^2}{2^{1-2 n} \pi  (n+2) (n+3)} \frac{e^{2 \pi  \xi }}{\xi ^5 t^4}, & -2<n<-\frac{1}{2}
 \end{align}
 For $n>1$ or $n<-2$ we have an infrared divergence, specifically for $n=-2,1$ the integral diverges logarthmically. We therefore limit the discussion from now on to $-2<n<1$.
 Except for $n= 1,\,-2$ the expressions can be simplified to:
 \begin{align}
\frac{1}{2}\langle \vec{\tilde E}^2+\vec{\tilde B}^2 \rangle &\simeq \frac{1}{4\pi^2} \frac{ n^2 \Gamma (2 n+1)^2}{2^{3+2 n} \pi  (n-1)}\frac{e^{2 \pi  \xi }}{\xi ^3
   t^4} =\frac{1}{4\pi^2} \frac{ n^4 \Gamma (2 n)^2}{2^{1+2 n} \pi  (1-n)}\frac{e^{2 \pi  \xi }}{\xi ^3
   t^4}\equiv D_1(n) \frac{e^{2 \pi  \xi }}{\xi ^3
   t^4} , & -\frac{1}{2}<n<1\\
  \frac{1}{2}\langle \vec{\tilde E}^2+\vec{\tilde B}^2 \rangle &\simeq \frac{1}{4\pi^2} \frac{  (n+1)^2 \Gamma (-2 n-1)^2}{2^{1-2 n}\pi  (n+2)}\frac{e^{2 \pi  \xi }}{ \xi
   ^3 t^4}\equiv D_2(n)\frac{e^{2 \pi  \xi }}{\xi ^3
   t^4} , & -2<n<-\frac{1}{2}
   \end{align}
   As such, the bound from the Friedmann equation
   \be
   \label{boundII}
   H/M_{pl}\ll \sqrt{3/D_{1,2}(n)} \,p^2\xi^{3/2}e^{-\pi \xi}
   \ee
    has the same functional dependence as in model I, the only difference being some numerical factor $D_{1,2}(n)$ that depends on $n$. Actually, since we know about the symmetry of $n\rightarrow -1-n, \gamma \rightarrow -\gamma n/(1+n)$ we could have simply performed the transformation on $D_1$ without explicitly calculating $D_2$ and assess the backreaction constraint accordingly.
   The bound from the field equation will be:
   \be
\Bigg | \frac{2 n}{a_1}\left(\frac{1}{2}\langle \vec{\tilde E}^2+\vec{\tilde B}^2 \rangle-\gamma\langle \vec{\tilde E}\cdot \vec{\tilde B}\rangle\right) \Bigg |\ll |\ddot{\varphi_1}|\sim|V'|
  \ee
 Since the strongest bound in terms of the small parameter $p \ll 1$ is given by the $F^2$ term, we can neglect the $F\tilde F$ contribution. Using $a_1\sim 1/\sqrt{2p}$ gives:
 \be
 \frac{H}{M_{p}}\ll \frac{1}{\sqrt{2n D_{1,2}(n)}}p\xi^{3/2}e^{-\pi \xi}
 \ee 
 that is parametrically easier to fulfil compared to bound coming from the Friedmann equation, as in model I.

 To summarize, in both models: If at some very small $t<0$ before the bounce the gauge field production shuts down (presumably during kinetic contraction).  Then i) It is possible to have a coupling between ekpyrosis and gauge fields. ii) It is possible to exponentially enhance the gauge fields. Since in realistic models there is a turn in field space, one may hope that the shutting down mechanism of gauge field production could be part of this turn, such that no new ingredients have to be added to the model, and the mechanism described here is not destroyed.
  
  \section{Formal Solutions, Correlators and the Green's Function}
\label{sec:formal-sol}

We are now in a position to calculate the tensor spectra generated due to the source term in the field equations, rather 
than the standard vacuum fluctuations. We use the formalism laid out in \cite{Barnaby:2012xt}.

Decomposing the spatial part of the metric $g_{ij} = a^2 \left( \delta_{ij} + h_{ij} \right)$, where the modes $h_{ij}$ are transverse and traceless. Introducing the canonical modes:
\begin{equation}
 \frac{M_p a}{2} h_{ij} =  \int \frac{d^3 k}{\left( 2 \pi \right)^{3/2}} \, {\rm e}^{i \vec{k} \cdot \vec{x}}  \, \sum_{ \lambda =  \pm } \, \Pi_{ij,\lambda} \left( {\hat k} \right) \,  Q_\lambda \left( \vec{k} \right) \;\;\;,\;\;\;
\Pi_{ij,\lambda} \left( {\hat k} \right) = \epsilon^{(\lambda)}_i \left( {\hat k} \right)  \epsilon^{(\lambda)}_j \left( {\hat k} \right) 
\label{formal-hc-def}
\end{equation}
which  obey the  equation
\begin{equation}
\left[ \partial_\tau^2 + \left( k^2 - \frac{a''}{a} \right) \right]  Q_\lambda \left( \tau ,\, \vec{k} \right) = J_\lambda \left( \tau ,\, \vec{k} \right)
\label{hc-eq-formal}
\end{equation}

The source is obtained by considering the transverse and traceless spatial part of the energy momentum tensor and projecting along the $\lambda$ polarization with the polarization tensor $\Pi_{ij,\lambda}$: 
\begin{equation}
J_\lambda \left( \tau ,\, \vec{k} \right) = \Pi_{ij,\lambda}^{*} \left( {\hat k} \right) \int \frac{d^3 x}{\left( 2 \pi \right)^{3/2}} \, {\rm e}^{-i \vec{k} \cdot \vec{x}} \, \frac{a}{M_p} \, T_{ij} \left( \tau ,\, \vec{x} \right)
\label{Jlambda-formal}
\end{equation}
%
If the energy-momentum tensor is quadratic in the gauge fields as in our case, we can write a formal expression of the source term in terms of an operator
 $ {\hat O}_{\lambda,ij} \left( \tau ,\, \vec{k} ,\, \vec{p} \right) $:

%

\begin{eqnarray}
J_{\lambda}  \left( \tau ,\, \vec{k} \right) \equiv  \int \frac{ d^3 p }{ \left( 2 \pi \right)^{3/2} } \; {\hat O}_{\lambda,ij} \left( \tau ,\, \vec{k} ,\, \vec{p} \right) \, {\tilde A_i} \left( \tau ,\, \vec{p} \right)  {\tilde A_j} \left( \tau ,\, \vec{k} - \vec{p} \right)  \;\;\;,\;\;\; \left\{ \lambda=\pm \right\}\cr
\label{pbm-formal}
\end{eqnarray}
%
$ {\hat O}_{\lambda,ij}$ is invariant under simultaneous $i \leftrightarrow j$ and $\vec{p} \rightarrow \vec{k} - \vec{p}$ operations.


The equation (\ref{hc-eq-formal}) is formally solved by
\begin{equation}
Q_{\lambda} \left( \tau ,\, \vec{k} \right) =  Q_{\lambda,{\rm v}}  \left( \tau ,\, \vec{k} \right) 
+  Q_{\lambda,{\rm s}}  \left( \tau ,\, \vec{k} \right) 
\label{formal-sol1}
\end{equation}
where $ Q_{\lambda,{\rm v}}  $ is the standard vacuum solution of the homogeneous equation. The sourced term is the particular solution given by:
\begin{eqnarray}
 Q_{\lambda,{\rm s}}  \left( \tau ,\, \vec{k} \right) & = &  \int^\tau d \tau' G_k \left( \tau ,\, \tau' \right) \, J_\lambda \left( \tau' ,\, \vec{k} \right) \nonumber\\
\label{formal-sol2}
\end{eqnarray}
where $G_k \left( \tau ,\, \tau' \right)$ is the Green function, that we have to calculate in the relevant background.
To reduce clutter, we write the scale factor as $a(\tau)\sim (-\tau)^b$, so according to our previous conventions $b\equiv p/(1-p)\sim p$.
Let us note, that once we find the Green function, it is valid for other FLRW backgrounds such as inflationary $(b=-1)$, and matter bounce $b=2$.
We are interested in the retarded Green function that solves:
\be
G''+\left(k^2-\frac{a''}{a}\right)G=\delta(\tau-\tau')
\ee
with boundary conditions $G(\tau'>\tau)=0$.
The general solution to the homogeneous equation is:
\be
G=i \frac{\sqrt{\pi}}{2}\sqrt{-\tau}H_{\nu}^{(1)}(-k\tau),\quad \nu=\frac{1}{2}-b
\ee
Imposing the boundary condition, and a discontinuity in the derivative, gives, up to an irrelevant phase:
\be
G_{ret.}(\tau, \tau')=i\Theta(\tau-\tau')\frac{\pi}{4}\sqrt{\tau \tau'}\left[H_{1/2-b}^{(1)}(-k\tau)H_{1/2-b}^{(2)}(-k\tau')-H_{1/2-b}^{(1)}(-k\tau')H_{1/2-b}^{(2)}(-k\tau)\right]
\ee
 
 Expanding outside the horizon $ - k \, \tau \ll 1$, 
 \be
 G_{ret.}(\tau, \tau')=\Theta(\tau-\tau')\sqrt{\tau \tau'}\frac{\Gamma(\nu)2^{\nu-2}}{(-k\tau)^{\nu}}2J_{\nu}(-k \tau')=
  \frac{\Theta(\tau-\tau')\Gamma(1/2-b)}{2^{1/2+b}k}(-k \tau)^b\sqrt{-k\tau'}J_{1/2-b}(-k\tau')
 \ee
Expanding near $b=0$ to zeroth order gives:
 \be
G_{ret.}(\tau,\tau') \simeq\Theta(\tau-\tau')\frac{\sin(-k\tau')}{k}
\ee

 In the cyclic/ekpyrotic scenario, the ekpyrotic phase is followed by a kinetic (energy) \textit{contracting phase}, then the bounce occurs, another kinetic \textit{expanding} phase, and then radiation domination.
In the kinetic phases $b=1/2$, the vacuum solutions involve $H_0^{(1)}(-k\tau), H_0^{(2)}(-k\tau)$. However, evaluating the Green's function for $b=1/2$, one derives:
\be
G_{ret.}(-k \tau \ll1)\simeq \sqrt{-k\tau}\cdots \rightarrow_{-k \tau\rightarrow 0} 0
\ee
So even in the presence of gauge fields, gravitational waves are not produced from the interaction with gauge fields during the kinetic energy domination phase. Therefore, the last mode to be generated and exit the horizon, will be the mode at the end of ekpyrosis $k=H_{end}$.

It is important to note that the vacuum solution and the sourced one (or equivalently the homogeneous and particular solution) are statistically independent of one another. This is because the vacuum solution is made up of the $b,b^{\dagger}$ operators, while the sourced solution is made up from the gauge field operators $a,a^{\dagger}$ and from \eqref{abcommute}, the $a,b$ operators commute. Therefore, there is no cross correlation between the two contributions and each correlator will be a sum of the different autocorrelators. For example, the power spectrum will be:
\be
{\cal P}_T^{total}={\cal P}_T^v+{\cal P}_T^s,
\ee
where we calculated ${\cal P}_T^v$ in \eqref{PTv}, we now turn to the calculation of ${\cal P}_T^s$.
\section{Sourced GW Spectrum Model I}
\label{sec:model1}
Writing again the general form of the tensor perturbation and it decomposition to $\pm$ polarization:
\bea
\hat{h}_{ij}=\frac{2}{M_{pl}a}\int \frac{d^3k}{(2\pi)^{3/2}}e^{i \vec{k}\vec{x}}\sum_{\lambda=\pm}\Pi^*_{ij,\lambda}(\hat k)\hat Q_{\lambda}(\tau ,\vec{k})\\
\hat{h}_{\lambda}(\tau,\vec{k})\equiv \Pi_{ij,\lambda}\hat{h}_{ij}(\tau,\vec{k})=\frac{2}{M_{pl}a}\hat{Q}_{\lambda}(\tau,\vec{k})
\eea
Some useful formulae in Fourier space of the source term for tensors and the gauge fields are:
\bea
J_{\lambda}(\tau, \vec{k})&=&-\frac{a^3}{M_{pl}}\Pi_{ij,\lambda}(\hat k)\int \frac{d^3x}{(2\pi)^{3/2}}e^{-i \vec{k}\vec{x}}\left[\hat E_i \hat E_j+\hat B_i \hat B_j\right](x^i)\\
\hat A_i(\tau, \vec{k})&=&\sum_{\lambda=\pm}\epsilon_i^{(\lambda)}(\hat k)\tilde A_{\lambda}(\tau,k)\left[\hat a_{\lambda}(\vec k)+\hat{a}_{\lambda}^{\dagger}(-\vec k)\right]\\
\hat E_i^{(\lambda)}&=&-\frac{1}{a^2}\epsilon_i^{(\lambda)}(\hat k)\sqrt{\frac{2 k \xi}{-\tau}}\tilde A_{\lambda}(\tau,k)\left[\hat a_{\lambda}(\vec k)+\hat{a}_{\lambda}^{\dagger}(-\vec k)\right]\\
\hat B_i^{(\lambda)}&=&\frac{1}{a^2}\epsilon_i^{(\lambda)}(\hat k)\lambda k \tilde A_{\lambda}(\tau,k)\left[\hat a_{\lambda}(\vec k)+\hat{a}_{\lambda}^{\dagger}(-\vec k)\right]\\
 \tilde A_{+}(\tau,k) &\simeq& \frac{1}{\sqrt{2k}} \left(\frac{-k\tau}{2\xi}\right)^{1/4} e^{\pi \xi - 2\sqrt{-2\xi k\tau}}
 \eea
  Substituting the above equations and the definition of $\Pi_{ij,\lambda}$ gives:
  \bea
J_{\lambda}(\tau, \vec{k})  =-\frac{1}{M_{pl}a}\int \frac{d^3p}{(2\pi)^{3/2}}\sum_{\lambda'=\pm}\epsilon_i^{(\lambda)*}(\vec k)\epsilon_j^{(\lambda)*}(\vec k)\epsilon_i^{\lambda'}(\vec p)\epsilon_j^{\lambda'}(\vec k -\vec p)\left\{\frac{2 \xi}{-\tau}\sqrt{p|\vec k-\vec p|}+p|\vec k-\vec p|\right\}\cr
\times\tilde A_{\lambda'}(\tau, \vec p)\tilde A_{\lambda'}(\tau,\vec k-\vec p)\left[\hat a_{\lambda'}(\vec p)+\hat{a}_{\lambda'}^{\dagger}(-\vec p)\right]
\left[\hat a_{\lambda'}(\vec k-\vec p)+\hat{a}_{\lambda'}^{\dagger}(-\vec k+\vec p)\right]\cr
\eea
We are interested only in the $+$ polarization that enhances the gauge field fluctuation.
Substituting the source term into the gravitational wave operator $\hat h_{\lambda}$ gives:
\bea
\hat h_{\lambda}=-\frac{2}{M_{pl}^2a(\tau)}\int^{\tau}d\tau'\frac{G_k(\tau,\tau')}{a(\tau')}\int \frac{d^3p}{(2\pi)^{3/2}}P_{\lambda}(\vec k, \vec p,\vec k-\vec p)
\left\{\frac{2 \xi}{-\tau'}\sqrt{p|\vec k-\vec p|}+p|\vec k-\vec p|\right\}\cr
\times\tilde A_{\lambda'}(\tau', \vec p)\tilde A_{\lambda'}(\tau',\vec k-\vec p)\left[\hat a_{\lambda'}(\vec p)+\hat{a}_{\lambda'}^{\dagger}(-\vec p)\right]
\left[\hat a_{\lambda'}(\vec k-\vec p)+\hat{a}_{\lambda'}^{\dagger}(-\vec k+\vec p)\right]
\eea
with 
\be
P_{\lambda}(\vec k, \vec p,\vec k-\vec p)=\epsilon_i^{(\lambda)*}(\vec k)\epsilon_i^{+}(\vec p)\epsilon_j^{(\lambda)*}(\vec k)\epsilon_j^{+}(\vec k -\vec p)
\ee
Substituting $\tilde A$ we arrive at:
\bea
\hat h_{\lambda}&=&-\frac{2e^{2\pi \xi}}{M_{pl}^2k_e^{2b}}\int^{\tau}d\tau'\frac{G_k(\tau,\tau')}{(\tau \tau')^b}\int \frac{d^3p}{(2\pi)^{3/2}}P_{\lambda}(\vec k, \vec p,\vec k-\vec p)
\left\{\frac{2 \xi}{-\tau'}\sqrt{p|\vec k-\vec p|}+p|\vec k-\vec p|\right\}\cr
&\times&\frac{(p|\vec k-\vec p|)^{-1/4}}{2}\sqrt{\frac{-\tau'}{2\xi}}e^{-2\sqrt{-2\xi \tau'}\left(\sqrt{p}+\sqrt{|\vec k-\vec p|}\right)}\left[\hat a_{\lambda'}(\vec p)+\hat{a}_{\lambda'}^{\dagger}(-\vec p)\right]
\left[\hat a_{\lambda'}(\vec k-\vec p)+\hat{a}_{\lambda'}^{\dagger}(-\vec k+\vec p)\right]\cr
&=&-\frac{e^{2\pi \xi}}{M_{pl}^2k_e^{2b}}\int^{\tau}d\tau'\frac{G_k(\tau,\tau')}{(\tau \tau')^b}\int \frac{d^3p}{(2\pi)^{3/2}}P_{\lambda}(\vec k, \vec p,\vec k-\vec p)
\left\{\frac{2 \xi}{-\tau'}+\sqrt{p|\vec k-\vec p|}\right\}\cr
&\times&(p|\vec k-\vec p|)^{1/4}\sqrt{\frac{-\tau'}{2\xi}}e^{-2\sqrt{-2\xi \tau'}\left(\sqrt{p}+\sqrt{|\vec k-\vec p|}\right)}\left[\hat a_{\lambda'}(\vec p)+\hat{a}_{\lambda'}^{\dagger}(-\vec p)\right]
\left[\hat a_{\lambda'}(\vec k-\vec p)+\hat{a}_{\lambda'}^{\dagger}(-\vec k+\vec p)\right]\cr
&\simeq&-\frac{e^{2\pi \xi}\Gamma(1/2-b)\sqrt{\xi}k^{b-1/2}}{2^bM_{pl}^2k_e^{2b}}\int^{\tau}d\tau'\frac{J_{1/2-b}(-k \tau')}{(-\tau')^b}\int \frac{d^3p}{(2\pi)^{3/2}}P_{\lambda}(\vec k, \vec p,\vec k-\vec p)\cr
&\times&(p|\vec k-\vec p|)^{1/4}e^{-2\sqrt{-2\xi \tau'}\left(\sqrt{p}+\sqrt{|\vec k-\vec p|}\right)}\left[\hat a_{\lambda'}(\vec p)+\hat{a}_{\lambda'}^{\dagger}(-\vec p)\right]
\left[\hat a_{\lambda'}(\vec k-\vec p)+\hat{a}_{\lambda'}^{\dagger}(-\vec k+\vec p)\right]\cr
\eea
where $\simeq$ is because we are interested only in the dominant contribution, outside the horizon, when $-k\tau' \ll 1$.
Let us perform a change of variables to $x=-k \tau$ and do the time integral. 
\begin{align}
\hat h_{\lambda}&\simeq-\frac{e^{2\pi \xi}\Gamma(1/2-b)\sqrt{\xi}k^{2b-3/2}}{2^bM_{pl}^2k_e^{2b}}\int \frac{d^3p}{(2\pi)^{3/2}}\int_{x_{end}}^{\infty}dx' \frac{J_{1/2-b}(x')}{x'^b}e^{-2\sqrt{-2\xi x'/k}\left(\sqrt{p}+\sqrt{|\vec k-\vec p|}\right)}\cr
&\times P_{\lambda}(\vec k, \vec p,\vec k-\vec p)
(p|\vec k-\vec p|)^{1/4}\left[\hat a_{\lambda'}(\vec p)+\hat{a}_{\lambda'}^{\dagger}(-\vec p)\right]
\left[\hat a_{\lambda'}(\vec k-\vec p)+\hat{a}_{\lambda'}^{\dagger}(-\vec k+\vec p)\right]
\end{align}
Performing the $x'$ integration with $x_{end}\rightarrow 0$ and $z=2\sqrt{-2\xi/k}\left(\sqrt{p}+\sqrt{|\vec k-\vec p|}\right)$, and everywhere except in $(k/k_e)^{2b}$ taking $b\rightarrow 0$:
\be
\int_0^{\infty}dx'J_{1/2}(x')e^{-z\sqrt{x'}}=\sqrt{\frac{2}{\pi}}\frac{4}{z^3}
\ee
\begin{align}
\hat h_{\lambda}&\simeq-\frac{e^{2\pi \xi}k^{2b-3/2}}{4\xi M_{pl}^2k_e^{2b}}\int \frac{d^3p}{(2\pi)^{3/2}} P_{\lambda}(\vec k, \vec p,\vec k-\vec p)
\frac{(p|\vec k-\vec p|)^{1/4}}{k^{-3/2}\left(\sqrt{p}+\sqrt{|\vec k-\vec p|}\right)^3} \cr
&\times \left[\hat a_{\lambda'}(\vec p)+\hat{a}_{\lambda'}^{\dagger}(-\vec p)\right]
\left[\hat a_{\lambda'}(\vec k-\vec p)+\hat{a}_{\lambda'}^{\dagger}(-\vec k+\vec p)\right]\
\end{align}

The dimensionless sourced tensor power spectrum is defined according to:
\be
\mathcal P_{\lambda}(k)\delta_{\lambda \lambda'}\delta(\vec k+\vec k')\equiv \frac{k^3}{2\pi^2}\Big \langle \hat h_{\lambda}(\vec k)\hat h_{\lambda'}(\vec k')\Big \rangle
\ee
Omitting the delta functions for brevity, the two point function then gives:
\bea
\Big \langle \hat h_{\lambda}(\vec k)\hat h_{\lambda'}(\vec k')\Big \rangle=2\times\frac{e^{4\pi \xi} k^{4b}}{16\xi^2 M_{pl}^4k_e^{4b}}
\int \frac{d^3p}{(2\pi)^3} |P_{\lambda}(\vec k, \vec p, \vec k-\vec p)|^2 \frac{\sqrt{p|\vec k-\vec p|}}
{\left(\sqrt{p}+\sqrt{|\vec k-\vec p|}\right)^6}
\eea

The factor of $2$ is due to the ladder operators, the rest is self-evident.
Choosing $\theta$ as the angle between $\vec k$ and $\vec p$ and switching to dimensionless momentum $\vec p=|k|\vec q$ and noting
\bea
\label{absP}
|P_{\lambda}(\vec k, \vec p, \vec k-\vec p)|^2=\frac{1}{16}\left(1+\lambda \frac{\vec k \cdot \vec p}{kp}\right)^2 \left(1+\lambda \frac{k^2-\vec k \cdot \vec p}{k|\vec k-\vec p|}\right)^2\cr
=\frac{(1\pm \cos \theta)^2\left(1-q\cos \theta\pm\sqrt{1-2q \cos \theta+q^2}\right)^2}{16(1-2q \cos \theta+q^2)},
\eea
we arrive at the following expression:
\be
\Big \langle \hat h_{\lambda}(\vec k)\hat h_{\lambda'}(\vec k')\Big \rangle=\frac{e^{4\pi \xi} k^{1+4b}}{2^7\xi^2 M_{pl}^4k_e^{4b}}\int \frac{d \cos \theta dq q^2}{(2\pi)^2} \sqrt{q|\hat k-\vec q|}\frac{\left(1\pm \cos \theta\right)^2 \left(1-q\cos \theta\pm\sqrt{1-2q \cos \theta+q^2}\right)^2}{(1-2q \cos \theta+q^2)\left(\sqrt{q}+\sqrt{|\hat k-\vec q|}\right)^6}
\ee

Since the momentum integral is diverging linearly, it will be dominated by the UV cut-off, $H_{end}$. Hence, we can consider the contribution of the large $q \gg1$ limit.
In such a case, the integral can be performed analytically in a simple way:
\bea
\int \frac{d \cos \theta dq q^2}{(2\pi)^2} \sqrt{q|\hat k-\vec q|}\frac{\left(1\pm \cos \theta\right)^2 \left(1-q\cos \theta\pm\sqrt{1-2q \cos \theta+q^2}\right)^2}{(1-2q \cos \theta+q^2)\left(\sqrt{q}+\sqrt{|\hat k-\vec q|}\right)^6}\cr
\simeq \int \frac{d \cos \theta dq}{(2\pi)^2} \left[\frac{(1-\cos^2\theta)^2}{64}+\mathcal{O}(q^{-1})\right]=\frac{q_{UV}}{30(2\pi)^2}=\frac{H_{end}}{30(2\pi)^2k}
\eea

So, the sourced tensor power spectrum produced by inverse decays of the gauge quanta, is given by:
\bea
{\cal P}_T^{s}
&=&
 \frac{e^{4\pi \xi}} {2^7\xi^2 30(2\pi)^2 2\pi^2} \frac{k^{3+4b}H_{end}}{M_{pl}^4k_e^{4b}}\\
{\cal P}_T^{total}&=&\frac{4k^{2+2b}}{\pi^2 M_{pl}^2k_e^{2b}}\left[1+\frac{e^{4\pi \xi}} {\xi^2}\frac{k^{1+2b} H_{end}}{M_{pl}^2k_e^{2b}}\frac{1}{2^{12}\pi^2\times 60}\right]=\frac{4k^{2+2b}}{\pi^2 M_{pl}^2k_e^{2b}}\left[1+8.2\times10^{-7}\, \frac{e^{4\pi \xi}} {\xi^2}\frac{k^{1+2b} H_{end}}{M_{pl}^2k_e^{2b}}\right]\cr
\eea
Substituting the bounds from the backreaction that will destroy the background evolution, and evaluating at $k_e=H_{end}/b, \quad b=1/50$, shows that the inverse decay dominate for $\xi\gtrsim 3.1$. However, as we can immediately see, the resulting spectrum is even more blue than the vacuum one. In the inflationary case ${\cal P}_{s}\sim {\cal P}_{vac}^2$ and naively this repeats itself in ekpyrosis. The main difference is that here we had a sharp cut-off $H_{end}$ instead of integrating to infinite wave-number $k$. Hence, instead of getting $k^{4+4b}$ dependence, we got, $H_{end} k^{3+4b}$, which is still bluer than the vacuum one, and as such, will still be unobservable at CMB scales. It will be interesting to check whether interesting bounds can be derived from the recent advanced LIGO results \cite{Abbott:2016blz}.

\section{Sourced GW Spectrum Model II}
\label{sec:model2}
The formal expression for the sourced GW is identical to model I, the only difference being the different mode functions for the photons. For the $+$ polarization: 
\bea
\label{hsorbo}
\hat h_{\lambda}&\simeq&-\frac{2}{M_{pl}^2a(\tau)}\int^{\tau}d\tau'\frac{G_k(\tau,\tau')}{a(\tau')}\int \frac{d^3p}{(2\pi)^{3/2}}P_{\lambda}(\vec k, \vec p,\vec k-\vec p)
\partial_{\tau'}\tilde A_{\lambda'}(\tau', \vec p)\partial_{\tau'}\tilde A_{\lambda'}(\tau',\vec k-\vec p)\cr
&\times&\left[\hat a_{\lambda'}(\vec p)+\hat{a}_{\lambda'}^{\dagger}(-\vec p)\right]
\left[\hat a_{\lambda'}(\vec k-\vec p)+\hat{a}_{\lambda'}^{\dagger}(-\vec k+\vec p)\right]
\eea
As we have seen, the parameter space for $n$ is divided between $1>n>-1/2$ and $-1/2>n>-2$, calling them regions $r=1,2$ respectively. Observables will be related in the two regions via the transformation $n\rightarrow -1-n$. \footnote{Except in the power of $k, k_e$, we will usually take $b\rightarrow 0$ limit for simplicity. Keeping the finite $b$ has negligible effect on the result.} We can write the general expression of the two point correlator:
\bea
\Big \langle \hat h_{\lambda}(\vec k)\hat h_{\lambda'}(\vec k')\Big \rangle&=&\delta_{\lambda \lambda'}\delta(\vec k+\vec k')\, 2\, {\cal N}_{r}^2{\cal I}_{r}^2\,\frac{e^{4\pi \xi}\xi^{2\alpha}}{M_{pl}^4}\,\int \frac{d^3p}{(2\pi)^3} |P_{\lambda}(\vec k, \vec p,\vec k-\vec p)|^2
\left( p
   |\vec k- \vec p|\right)^{\alpha}\\
  &=&\delta_{\lambda \lambda'}\delta(\vec k+\vec k')\, 2\,{\cal N}_{r}^2{\cal I}_{r}^2\,\frac{e^{4\pi \xi}\xi^{2\alpha}}{M_{pl}^4} \,k^{3+2\alpha}\int \frac{d \cos \theta dq q^2}{(2\pi)^2} |P_{\lambda}|^2
\left( q
   |\hat k- \vec q|\right)^{\alpha}\\
{\cal P}_T^s &=& \frac{{2\cal N}_{r}^2{\cal I}_{r}^2}{2\pi^2}\, \frac{e^{4\pi \xi}\xi^{2\alpha}}{M_{pl}^4}\, k^{6+2\alpha}\int \frac{d \cos \theta dq q^2}{(2\pi)^2} |P_{\lambda}(\hat k, \vec q,\hat k-\vec q)|^2
\left( q
   |\hat k- \vec q|\right)^{\alpha}
\eea
The factor of $2$ coming from the creation and annihilation operators, ${\cal N}_{r}$ denotes different numerical factors we shall later specify, ${\cal I}_{r}$ the time integral for each region and $\alpha=(-1)^r(2n+1)$, and we have switched to the dimensionless momentum $\vec p=|k|\vec q$. The power spectrum is symmetric around $n=-1/2$. Again due to the duality \eqref{eq:duality} we shall have 
\be
{\cal P}_T(n,\xi ;\, n\geq-1/2)={\cal P}_T(-1-n, \xi; \, n\leq -1/2).
\ee
Therefore, we shall simply calculate the spectra for $-1/2\geq n>-2$. We have verified that the spectra are identical by a full explicit independent calculation of both regions.
Notice that in certain regions of the parameter space, where the momentum integral is finite \textit{and} ${\cal I}$ independent of $k$, we can already read off the tilt of the correlator and the corresponding power spectrum. This will happen for $-2<n<-5/4$ and $1/4<n<1$. Therefore as $n\rightarrow -2$ or $n \rightarrow 1$ we will get a slightly blue power spectrum approaching \textit{scale invariance!}

\subsection{$-2<n<-1/2$} 

Substituting the $\tilde A$ for $n<-1/2$:
\bea
\label{modeh2}
\hat h_{\lambda}&\simeq&-\frac{2}{M_{pl}^2a(\tau)}\int^{\tau}d\tau'\frac{G_k(\tau,\tau')}{a(\tau')}\int \frac{d^3p}{(2\pi)^{3/2}}P_{\lambda}(\vec k, \vec p,\vec k-\vec p)
\frac{4^n e^{2 \pi  \xi } (n+1)^2 \xi^{1+2n}  \Gamma (-2 n-1)^2 }{\pi } (-\tau')^{2 n}\cr
&\times& 
 (p |\vec k- \vec p|)^{n+\frac{1}{2}} \left[\hat a_{\lambda'}(\vec p)+\hat{a}_{\lambda'}^{\dagger}(-\vec p)\right]
\left[\hat a_{\lambda'}(\vec k-\vec p)+\hat{a}_{\lambda'}^{\dagger}(-\vec k+\vec p)\right]\\
&\equiv&{\cal N}_2\frac{e^{2\pi \xi}\xi^{1+2n}}{M_{pl}^2}\int^{\tau}d\tau'\frac{G_k(\tau,\tau')}{a(\tau)a(\tau')}(-\tau')^{2 n}\int \frac{d^3p}{(2\pi)^{3/2}}P_{\lambda}(\vec k, \vec p,\vec k-\vec p)
 ( p |\vec k- \vec p|)^{n+\frac{1}{2}} [\hat a+ \hat a^{\dagger}][\hat a+ \hat a^{\dagger}]\cr
\eea
where we defined
\be
{\cal N}_2=\frac{-2 \times4^n (n+1)^2 \Gamma (-2 n-1)^2 }{\pi}
\ee
Notice a useful equality that will be helpful in evaluating the bounds from backreaction from section \ref{sec:backreaction}:
\be
{\cal N}_2=-4(n+2) 4\pi^2D_2
\ee
Consider the time integral for $x\equiv-k\tau \ll1$:
\begin{align}
\label{I2}
{\cal I}&\equiv \int^{\tau}d\tau'\frac{G_k(\tau,\tau')}{a(\tau)a(\tau')}(-\tau')^{2 n}\simeq \frac{k^{-2+2b-2n}}{k_e^{2b}}\left(\frac{\Gamma(1/2-b)\Gamma(1-b+n)}{2^{2b-2n}\Gamma(1/2-n)}-\frac{x_{end}^{2-2b+2n}}{2(1-b+n)}+{\cal O}(x_{end}^{4-2b+2n})\right)\cr
&=\frac{1}{k_e^{2b}}\left(\frac{\Gamma(1/2-b)\Gamma(1-b+n)}{2^{2b-2n}\Gamma(1/2-n)}k^{-2+2b-2n}-\frac{(-\tau_{end})^{2-2b+2n}}{2(1-b+n)}\right)
\equiv k_e^{-2b}\left(C_{02} k^{-2+2b-2n}+C_{12}(-\tau_{end})^{2-2b+2n}\right)
\end{align}
The first term in the brackets dominates if $-1<n<-1/2$ while the second for $-2<n<-1$, for $b\ll1$. The higher order terms are always negligible.
The rest of the section is devoted to calculation of the momentum integral and ${\cal I}$ for various $n$.
\subsubsection{$-1<n<-1/2$}

\begin{align}
\Big \langle \hat h_{\lambda}(\vec k)\hat h_{\lambda'}(\vec k')\Big \rangle &\simeq 2{\cal N}_2^2\times\frac{e^{4\pi \xi}\xi^{4n+2}}{M_{pl}^4}\times\frac{C_{02}^2k^{-4+4b-4n}}{k_e^{4b}} k^{5+4n}\cr
&\times\int \frac{d\cos \theta dq q^2}{(2\pi)^2}\frac{(1\pm \cos \theta)^2\left(1-q\cos \theta\pm\sqrt{1-2q \cos \theta+q^2}\right)^2}{16(1-2q \cos \theta+q^2)}
\left(q|\hat k-\vec q|\right)^{2n+1}\cr
&\simeq {\cal N}_2^2 \times \frac{e^{4\pi \xi}\xi^{4n+2}}{M_{pl}^4} \times \frac{C_{02}^2k^{1+4b}}{k_e^{4b}}\frac{q_{end}^{5+4n}}{15(5+4n)\pi^2}.
\end{align}
The above momentum integral is a good approximation as long as $n>-5/4$, which is obviously the case here.
Substituting the numerical factors back from \eqref{modeh2} and approximating $b=0$ everywhere expect in $k$ gives:
\begin{align}
\Big \langle \hat h_{\lambda}(\vec k)\hat h_{\lambda'}(\vec k')\Big \rangle &=\frac{4^{1+4n} (n+1)^4\Gamma (-2
   n-1)^4 \Gamma (n+1)^2}{15 \pi ^3 (4 n+5) \Gamma
   \left(\frac{1}{2}-n\right)^2}\times e^{4 \pi  \xi }\xi ^{4 n+2} \times \frac{k^{-4+4b-4n}H_{end}^{5+4n}}{M_{pl}^4k_e^{4b}} 
\end{align}

So the sourced power spectrum is:
\be
{\cal P}_T^s=\frac{2^{1+8n}  (n+1)^4\Gamma (-2
   n-1)^4 \Gamma (n+1)^2}{15 \pi ^5 (4 n+5) \Gamma
   \left(\frac{1}{2}-n\right)^2}\times e^{4 \pi  \xi }\xi ^{4 n+2} \times \frac{k^{-1+4b-4n}H_{end}^{5	+4n}}{M_{pl}^4k_e^{4b}} 
   \ee
Here we see that we still have an exponential enhancement via the $e^{4\pi \xi}$ factor on the one hand, and control over the tilt of the GW via $n$.
In this case, the "reddest" possible tilt corresponds to $n=-1/2$, gives $n_T\simeq 1+4b\sim 1$, this is better than the vacuum spectrum, that gives $n_T=2+2b$!  Nevertheless, considering the bounds from $N_{eff}$ of section \ref{sec:constraints}, if the sourced spectrum dominates, then for $n_T=1,\, k_{UV}/k_0=10^{20}$ we get $r<10^{-11}$. Despite \textit{twenty orders of magnitude} improvement, such $r$ is still completely unobservable on CMB scales.

Naively it may seem as if the amplitude is divergent for $n=-1/2$, but this is incorrect. The divergence comes from approximating the mode functions in \eqref{aapproxsm}, that are useful for generic $n$. In each case where there might be a divergence, like $n=-1/2,-1$ etc. we can take a proper limit of the mode functions for that specific $n$ and get a finite answer.

\subsection{$-2<n<-1$}
When $-2<n<-1$ the backreaction on the contracting background is still under control. However, until now, we simplified calculations by take the time integral limit $\tau_{end}\rightarrow 0$. When $-2<n<-1$ this simplification will cause the integral to diverge. Hence, in this case, we really need to evaluate the time integral up to a finite $\tau_{end}\neq 0$. 
This will correspond to the $C_{12}$ term dominating in \eqref{I2}. 
As long as $-5/4<n$ the momentum integral is unchanged.
\subsubsection{ $-5/4<n<-1$}
 \begin{align}
\Big \langle \hat h_{\lambda}(\vec k)\hat h_{\lambda'}(\vec k')\Big \rangle &\simeq 2{\cal N}_2^2\times \frac{e^{4\pi \xi}\xi^{4n+2}}{M_{pl}^4} \times \frac{C_{12}^2 (-\tau_{end})^{4-4b+4n}}{k_e^{4b}} k^{5+4n}\cr
&\times\int \frac{d\cos \theta dq q^2}{(2\pi)^2}\frac{(1\pm \cos \theta)^2\left(1-q\cos \theta\pm\sqrt{1-2q \cos \theta+q^2}\right)^2}{16(1-2q \cos \theta+q^2)}
\left(q|\hat k-\vec q|\right)^{2n+1}\cr
&= {\cal N}_2^2\times \frac{e^{4\pi \xi}\xi^{4n+2}}{M_{pl}^4} \times \frac{C_{12}^2(-\tau_{end})^{4-4b+4n}k^{5+4n}}{k_e^{4b}}\frac{q_{end}^{5+4n}}{15(5+4n)\pi^2}.
\end{align}
Since we normalized $a_{end}=1, k_e=(-\tau_{end})^{-1}$, so $k_e=H_{end}/b$, the simple expression for the sourced GW spectrum is:
\be
{\cal P}_{T}^s= \tilde {\cal N} \times e^{4 \pi  \xi }\xi ^{4 n+2} \times b^{4+4n}\frac{k^{3}H_{end}}{M_{pl}^4}, 
\ee
where $\tilde {\cal N}$ stands for the numerical factors we did not explicitly write.
Notice that here $n$ and $b$ drop out of the spectral tilt! The outcome is a blue spectrum $P_T^s\sim k^3$, bluer than the vacuum one, and as such unobservable on CMB scales.

\subsubsection{$-2<n<-5/4$}
For $-2<n<-5/4$ the momentum integral is finite and has to be calculated numerically. 
An excellent approximation for $\lambda=+$ is given by: \footnote{$\lambda=-$ is subdominant and we neglect its contribution.}
\be
\int \frac{d\cos \theta dq q^2}{(2\pi)^2}\frac{(1\pm \cos \theta)^2\left(1-q\cos \theta\pm\sqrt{1-2q \cos \theta+q^2}\right)^2}{16(1-2q \cos \theta+q^2)}
\left(q|\hat k-\vec q|\right)^{2n+1}\simeq \frac{11.1}{16(2\pi)^2(2+n)}
\ee
\begin{align}
\Big \langle \hat h_{\lambda}(\vec k)\hat h_{\lambda'}(\vec k')\Big \rangle 
=2 {\cal N}_2^2\times \frac{e^{4\pi \xi}\xi^{4n+2}}{M_{pl}^4} \times \frac{C_{12}^2(-\tau_{end})^{4-4b+4n}k^{5+4n}}{k_e^{4b}}\frac{11.1}{16(2\pi)^2(2+n)}\cr
\simeq 11.1 \times\frac{ 2^{4 n-5} (n+1)^2 \Gamma (-2 n-1)^4}{\pi ^4 (n+2)}\times e^{4\pi \xi}\xi^{2+4n}\times b^{4+4n}\frac{k^{5+4n}}{H_{end}^{4+4n}M_{pl}^4}\\
{\cal P}_T^s\simeq 11.1 \times \frac{2^{4 n-6} (n+1)^2 \Gamma (-2 n-1)^4}{\pi ^6 (n+2)}\times e^{4\pi \xi}\xi^{2+4n}\times b^{4+4n}\frac{k^{8+4n}}{H_{end}^{4+4n}M_{pl}^4} \label{pts2}
\end{align}

\begin{figure}[t]
\centerline{
\includegraphics{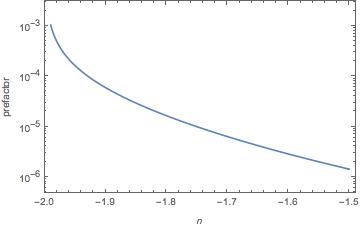}
}
\caption{Logplot of the prefactor in \eqref{pts2} as a function of $n$. 
}
\label{fig:prefactorn2}
\end{figure}

The numerical values of the prefactor in \eqref{pts2} are plotted in figure \ref{fig:prefactorn2}.
We see, that we got a finite GW spectrum, that is exponentially enhanced and that its tilt is controlled by $n$. For various $n$, all tilts between $0< n_T\leq 3$ are allowed. For example, $n=-15/8$ will correspond to $n_T=0.5$, which is still allowed by observations, while $n\rightarrow -2$ will correspond to a scale invariant spectrum, similar to inflation. 

Upon transforming $n\rightarrow -1-n, \gamma\rightarrow -\gamma n/(1+n)$, as to keep $\xi$ invariant we get the GW spectra in the case of $-1/2\leq n<1$. We have verified the results by full independent explicit calculation.
So similarly to the case of $n\rightarrow -2$, here when $n\rightarrow 1$ the spectrum becomes scale invariant.

To recap, we have shown that in a bouncing cosmology with gauge fields coupled as in model II any spectral tilt  $0< n_T\leq 3$ is allowed. Specifically, we get a spectrum that arbitrarily approaches scale invariance, though slightly blue for $n\rightarrow 1$ and $n\rightarrow -2$. Better yet, in the case of $n<0$, the gauge fields can be the $U(1)$ SM gauge group in the weak coupling regime. The relation between $n$ and $n_T$ is given by:
\bea
n_T&=4(1-n),  \quad 1>n>1/4\\
n_T&=4(2+n), \quad -2<n<-5/4
\eea
where again, the tilt, as being part of the spectrum is invariant under the symmetry. Assuming $r=10^{-4},\,n_T\leq0.63$ allows for $1>n\geq0.8425$ or $-2<n\leq-1.8425$. 
\section{Spectrum and Observations}
\label{sec:observations}
Let us combine the results of the previous section with the vacuum calculation and the bounds on backreaction.
\be 
{\cal P}_T^{total}={\cal P}_T^v+{\cal P}_T^s
\ee
Hence, for $n>1/4$ and $n<-5/4$ respectively:
\bea
\label{PfinalII1}
{\cal P}_T^{total}\simeq \frac{4k^2}{\pi^2 M_{pl}^2}\left[1+\frac{11.1 n^6 \Gamma (2 n)^4}{ 4^{2 n+4}(1-n)\pi ^4}  \times e^{4 \pi  \xi }/\xi ^{4 n+2} \times \frac{(H_{end}/b)^{4n}k^{2-4n}}{M_{pl}^2}\right]\\
{\cal P}_T^{total}\simeq \frac{4k^2}{\pi^2 M_{pl}^2}\left[1+\frac{11.1(n+1)^2 \Gamma (-2 n-1)^4}{ 4^{4-2n}(n+2)\pi ^4}\times e^{4\pi \xi}\xi^{2+4n}\times \frac{k^{6+4n}}{(H_{end}/b)^{4+4n}M_{pl}^2}\right]
\label{PfinalII2}
\eea
One can explicitly check that the transformation $n\rightarrow -1-n, \, \xi\rightarrow \xi$ is in tact and one spectrum is mapped to another and that they are identical.
The sourced spectra easily dominates over the vacuum one, since for $b=1/50$ the numerical factors are already greater than unity in all the relevant parameter space. The main question is whether the exponential $e^{4\pi \xi}$ enhancement is sufficient to overcome large suppressions coming form $M_{pl},H_{end}$, while still leaving the background evolution unchanged.
Let us rewrite the sourced power spectrum in terms of the tilt $n_T=4(2+n)$, in the case of $n<-1/2$. (an identical result occurs for $n_T=4(1-n)$ by $n\rightarrow -1-n$ for $n>-1/2$):
\be
\label{ptsexact}
{\cal P}_T^s=\frac{11.1 (4-n_T)^2 \Gamma
   \left(3-n_T/2\right)^4}{2^{16-n_T}\pi ^6 n_T}\frac{e^{4\pi \xi}}{b^{4-n_T}\xi^{6-n_T}}\left(\frac{H_{end}}{M_{pl}}\right)^4\left(\frac{k}{H_{end}}\right)^{n_T}, \quad n<-5/4
   \ee
   In the case of $n_T \ll1$ this can be further simplified to:
   \be
 {\cal P}_T^s\simeq  \frac{11.1}{256\pi^6 n_T}\frac{e^{4\pi \xi}}{b^{4}\xi^{6}}\left(\frac{H_{end}}{M_{pl}}\right)^4\left(\frac{k}{H_{end}}\right)^{n_T}
 \ee
 \be
{\cal P}_T^s \simeq \frac{282}{n_T}\times  \frac{e^{4\pi \xi}}{\xi^{6}}\left(\frac{H_{end}}{M_{pl}}\right)^4\left(\frac{k}{H_{end}}\right)^{n_T}
\label{ptfunctional}
 \ee
where in the second line we have used $b=1/50$. Hence, after long calculations we have a simple expression for a slightly blue GW power spectrum $n_T\gtrsim 0$ in terms of the tilt $n_T$, the Hubble parameter at the end of ekpyrosis $H_{end}$ and $\xi\equiv -n \gamma$, the coupling between the bouncer and the gauge fields.

\begin{figure}
\centerline{
\includegraphics{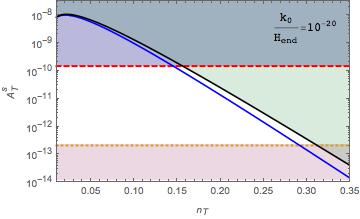}
}
\caption{The amplitude of the power spectrum at the pivot scale $A_T(k_0)=r A_s(k_0)=2.1\times 10^{-9}r$ as a function of $n_T$ for $k_0/H_{end}=10^{-20}$. The dashed red line corresponds to $r=0.07$. The shaded region above it is excluded by non-observation of GW by PLANCK+BICEP \cite{Array:2015xqh}. The dotted orange line corresponds to $r=10^{-4}$, below which, is undetectable in the next decade. The backreaction bound corresponds to the black solid line for $\xi=1$ and the blue solid line for $\xi=20$. Above these lines, the backreaction of the gauge fields is spoiling the background evolution, and the analysis will not be viable. We have a white trapezoid region of detectable GW on CMB scales $10^{-4}\leq r\leq 0.07$ with $n_T$ passing observational and backreaction constraints.
}
\label{fig:Atsnt20}
\end{figure}

\begin{figure}
\centerline{
\includegraphics{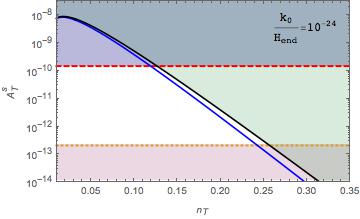}
}
\caption{ Same as figure \ref{fig:Atsnt20} but with $k_0/H_{end}=10^{-24}$
}
\label{fig:Atsnt24}
\end{figure}

For a simple evaluation of the backreaction constraint we reinsert $D_2$ into \eqref{ptsexact}: 
\bea
{\cal P}_T^s=\frac{11.1\times 4n_T}{(4-n_T)^2}D_2^2 \frac{e^{4\pi \xi}}{b^{4-n_T}\xi^{6-n_T}}\left(\frac{H_{end}}{M_{pl}}\right)^4\left(\frac{k}{H_{end}}\right)^{n_T}, 
\eea
Then, using equation \eqref{boundII} and $b\sim p$, the constraint on the spectrum is approximately:
\be
{\cal P}_T^s\ll \frac{33.3 \times 4n_T}{(4-n_T)^2}\xi^{n_T}b^{4+n_T}\left(\frac{k}{H_{end}}\right)^{n_T}
\ee
Taking $b=1/50$, 
the allowed region of parameter space is depicted in figures \ref{fig:Atsnt20},\ref{fig:Atsnt24}. 

We plot (Log plot) the amplitude of the power spectrum at the CMB pivot scale $A_T(k_0)=r A_s(k_0)=2.1\times 10^{-9}r$ as a function of $n_T$ for $k_0/H_{end}=10^{-20},10^{-24}$ in figures \ref{fig:Atsnt20},\ref{fig:Atsnt24} respectively. The dashed red line corresponds to $r=0.07$. The shaded region above it is excluded by non-observation of GW by PLANCK+BICEP \cite{Array:2015xqh}. The dotted orange line corresponds to $r=10^{-4}$, below which, is undetectable in the next decade. The backreaction bound corresponds to the black solid line for $\xi=1$ and the blue solid line for $\xi=20$. Above these lines, the backreaction of the gauge fields is spoiling the background evolution, and the analysis will not be viable. The bound from $N_{eff}$ is weaker than the backreaction constraints even in stage $4$ CMB future experiment, so it is not included in the plots. We see that there is a very weak dependence on $\xi$ and the main limitation comes from the value of $n_T$ and the number of wave modes between CMB scales $k_0$ and the end of ekpyrosis where GW production shuts down. 

In brief, the main outcome is the white trapezoid in Figures $4,5$, that depicts a detectable GW signal on CMB scales while passing all other theoretical and observational bounds and expressed by $n\leftrightarrow n_T, \, \xi, \,b$ and $H_{end}$. \footnote{We note that again, even considering scales all the way to the Planck scale, $k_0/H_{end}=10^{-30}$ there is still an allowed trapezoid region between $r=0.07, \, 0\leq n_T\leq0.09$ and $r=10^{-4},\, 0\leq n_T\leq 0.19$.}


   
\section{Final Remarks}
\label{sec:final}

In this work we have provided a proof of concept that observable GW on CMB scales can be generated in a bouncing cosmology which is not a matter bounce. The mechanism is based on the interaction of the bouncer field with gauge fields as suggested in Model II. The main results are given in \eqref{PfinalII1},\eqref{PfinalII2}, \eqref{ptfunctional} and in figures \ref{fig:Atsnt20} and \ref{fig:Atsnt24}.
We have also generalized earlier calculations, to be valid for general FLRW cosmologies in section \ref{sec:formal-sol}, and showed that Model I gives a very blue spectrum $n_T=3$, and hence unobservable on CMB scales.

If future CMB B-mode polarization observation is interpreted as a primordial GW signal, it does not provide a "proof" of inflation, since it might be that the signal is due to the mechanism presented here. There are two discriminators between inflation and our mechanism. First, a measurement of the helicity of the primordial signal, since the standard vacuum fluctuations predict a signal with both polarizations, while a tensor spectrum from the interaction with gauge fields has only one such polarization. Hopefully, such a conclusion can be reached in the case of a cosmic variance limited experiment \cite{Sorbo:2011rz}. Hence, if such a measurement is performed we will be able to answer decisively whether the GW spectrum is due to vacuum fluctuations or sourced fluctuations. However, that will not discern between a sourced spectrum of inflation or bouncing cosmology. Second, our mechanism predicts a \textit{blue tilt, $n_T\gtrsim 0$}, while slow-roll inflation, and most sourced inflationary scenarios predicts a red tilt $n_T\lesssim 0$. Thus, while both inflation and bouncing cosmologies can generate detectable GW signal on CMB scales, we shall have a definite answer after measuring the tilt and helicity of the spectrum.

Our suggestion has passed the immediate tests of maintaining the background evolution and having a viable GW amplitude and tilt once we assume the observed scalar spectrum. We have also demonstrated that the GW production stops once the Universe becomes kinetic energy dominated as it approaches the bounce. It is also in accord with current bounds coming from $N_{eff}$ as well as future bounds of $\Delta N_{eff}=0.02$. \footnote{We note again that it seems our bounds from $N_{eff}$ are somewhat conservative since the largest wave numbers will enter the horizon during kinetic domination. The tilt of these modes will be reduced $n_T\rightarrow n_T-1$, during kinetic energy domination, and therefore will yield weaker bounds from $N_{eff}$, after accounting for this phenomena.}

However, this work in itself cannot be considered a full model, since we have not accounted for the observed scalar power spectrum, the Universe sailing through the bounce in the presence of gauge fields and non-gaussianity bounds.
While a proper account of each of these aspects might spoil the above results, we believe that a full fledged model of a bouncing cosmology with observable GW on CMB scales is now in sight. 

{\bf \large Acknowledgements} \\
I thank Jean-Luc Lehners, Marco Peloso and Lorenzo Sorbo for invaluable discussions and correspondence.


\end{document}